\documentclass[graybox]{svmult}
\usepackage{mathptmx}       
\usepackage{helvet}         
\usepackage{courier}        
\usepackage{type1cm}        
\usepackage[bottom]{footmisc}
\usepackage{amsmath,amssymb}
\usepackage{mathbbol}
\usepackage[makeroom]{cancel}
\usepackage{verbatim}
\def\onehalf{{\textstyle\frac{1}{2}}}
\def\quarter{{\textstyle\frac{1}{4}}}
\def\HC{\mathcal{H}}
\def\LC{\mathcal{L}}

\def\bx{x}
\def\bj{\pmb{j}}
\def\ba{\pmb{a}}
\def\bp{\pmb{p}}
\def\bA{\pmb{A}}
\def\bF{\pmb{F}}
\def\bP{\pmb{P}}
\def\bPhi{\pmb{\Phi}}
\def\bphi{\pmb{\phi}}
\def\bPi{\pmb{\Pi}}
\def\bpi{\pmb{\pi}}
\def\RB{\mathbb{R}}
\def\bpartial{\pmb{\partial}}
\newcommand{\pfrac}[2]{\frac{\partial #1}{\partial #2}}
\begin{document}
\def\hbar{\mathchar'26\mkern-7mu h}
\title*{General U$(N)$ gauge transformations in the
realm of covariant Hamiltonian field theory}
\author{J\"urgen Struckmeier and Hermine Reichau}
\institute{J\"urgen Struckmeier \at Frankfurt Institute for Advanced Studies (FIAS)
\newline
Ruth-Moufang-Str.~1, 60438~Frankfurt am Main, Germany\newline
\email{struckmeier@fias.uni-frankfurt.de}
}
\maketitle
\abstract{
A consistent, local coordinate formulation of covariant Hamiltonian field theory is presented.
For regular Legendre transformations, the covariant canonical field equations are equivalent to the Euler-Lagrange field equations.
The covariant canonical transformation theory provides a systematic means of defining admissible mappings that preserve the variational principle --- and hence the form of the field equations --- with the corresponding transformation of the Hamiltonian.
Similar to the well-known canonical transformation theory of point dynamics, the canonical transformation rules for fields are derived from generating functions.
As an interesting example, we work out the generating function of type $F_{2}$ of a general local U$(N)$ gauge transformation and thus derive a {\em form-invariant\/} Hamiltonian description under {\em local\/} U$(N)$ gauge transformations, choosing the conventional quadratic gauge-field kinetic term.
For a particular regular Dirac Lagrangian, the resulting gauge-invariant Lagrangian $\LC_{3}$ includes Pauli-coupling of an $N$-tuple of fermions with the set of bosonic gauge fields.
Its coupling is controlled by an independent constant mass scale, and its relation to the Dirac-Pauli Lagrangian is a total divergence.
}
\section{\label{sec:d-w-theory}Covariant Hamiltonian density}
In field theory, the usual definition of a Hamiltonian density emerges from a Legendre transformation of a Lagrangian density $\LC$ that only maps the time derivative $\partial_{t}\phi$ of a field $\phi(t,x,y,z)$ into a corresponding canonical momentum variable, $\pi_{t}$.
Taking then the spatial integrals, we obtain a description of the field dynamics that corresponds to that of point dynamics.
In contrast, a fully covariant Hamiltonian description treats space and time variables on equal footing\cite{dedonder,weyl}.
If $\LC$ is a Lorentz scalar, this property is passed to the {\em covariant Hamiltonian}.
Moreover, this description enables us to derive a consistent theory of canonical transformations in the realm of classical field theory.
\subsection{\label{sec:kfgln}Covariant canonical field equations}
The transition from particle dynamics to the dynamics of a {\em continuous\/} system is based on the assumption that a {\em continuum limit\/} exists for the given physical problem\cite{saletan}.
This limit is defined by letting the number of particles involved in the system increase over all bounds while letting their masses and distances go to zero.
In this limit, the information on the location of individual particles is replaced by the {\em value\/} of a smooth function $\phi(\bx)$ that is given at a spatial location $x^{1},x^{2},x^{3}$ at time $t\equiv x^{0}$.
The differentiable function $\phi(\bx)$ is called a {\em field}.
In this notation, the index $\mu$ runs from $0$ to $3$, hence distinguishes the four independent variables of space-time $x^{\mu}\equiv(x^{0},x^{1},x^{2},x^{3})\equiv(t,x,y,z)$, and $x_{\mu}\equiv(x_{0},x_{1},x_{2},x_{3})\equiv(t,-x,-y,-z)$.
We furthermore assume that the given physical problem can be described in terms of a set of $I=1,\ldots,N$ --- possibly interacting --- scalar fields $\phi_{I}(\bx)$ or vector fields $\bA_{I}=(A_{I}^{0},A_{I}^{1},A_{I}^{2},A_{I}^{3})$, with the index ``$I$'' enumerating the individual fields.
In order to clearly distinguish scalar quantities from vector quantities, we denote the latter with boldface letters.
Throughout the article, the summation convention is used.
Whenever no confusion can arise, we omit the indexes in the argument list of functions in order to avoid the number of indexes to proliferate.

The Lagrangian description of the dynamics of a continuous system is based on the Lagrangian density function $\LC$ that is supposed to carry the complete information on the given physical system.
In a first-order field theory, the Lagrangian density $\LC$ is defined to depend on the $\phi_{I}$, possibly on the vector of independent variables $\bx$, and on the four first derivatives of the fields $\phi_{I}$ with respect to the independent variables, i.e., on the $1$-forms (covectors)
\begin{equation*}
\bpartial\phi_{I}\equiv(\partial_{t}\phi_{I},\partial_{x}
\phi_{I},\partial_{y}\phi_{I},\partial_{z}\phi_{I}).
\end{equation*}
The Euler-Lagrange field equations are then obtained as the zero of the variation $\delta S$ of the action integral
\begin{equation}\label{action-int}
S=\int\LC(\phi_{I},\bpartial\phi_{I},\bx)\,d^{4}x
\end{equation}
as\cite{saletan}
\begin{equation}\label{elgl}
\pfrac{}{x^{\alpha}}\pfrac{\LC}{(\partial_{\alpha}\phi_{I})}-
\pfrac{\LC}{\phi_{I}}=0.
\end{equation}
To derive the equivalent {\em covariant\/} Hamiltonian description of continuum dynamics, we first define for each field $\phi_{I}(\bx)$ a $4$-vector of conjugate momentum fields $\pi_{I}^{\mu}(\bx)$.
Its components are given by
\begin{equation}\label{p-def}
\pi_{I}^{\mu}=\pfrac{\LC}{(\partial_{\mu}\phi_{I})}
\equiv\pfrac{\LC}{\left(\pfrac{\phi_{I}}{x^{\mu}}\right)}.
\end{equation}
The $4$-vector $\bpi_{I}$ is thus induced by the Lagrangian $\LC$ as the {\em dual counterpart\/} of the $1$-form $\bpartial\phi_{I}$.
For the entire set of $N$ scalar fields $\phi_{I}(\bx)$, this establishes a set of $N$ conjugate $4$-vector fields.
With this definition of the $4$-vectors of canonical momenta $\bpi_{I}(\bx)$, we can now define the Hamiltonian density $\HC(\phi_{I},\bpi_{I},\bx)$ as the covariant Legendre transform of the Lagrangian density $\LC(\phi_{I},\bpartial\phi_{I},\bx)$
\begin{equation}\label{H-def}
\HC(\phi_{I},\bpi_{I},\bx)=\pi_{J}^{\alpha}
\pfrac{\phi_{J}}{x^{\alpha}}-\LC(\phi_{I},\bpartial\phi_{I},\bx).
\end{equation}
In order for the Hamiltonian $\HC$ to be valid, we must require the Legendre transformation to be {\em regular}.
This means that the full Hessian with respect to all field derivatives, with combined field and space-time indices, is non-singular, so that
\begin{equation*}
W_{I\mu,J\nu}=\pfrac{^{2}\LC}
{(\partial_{\mu}\phi_{I})\partial(\partial_{\nu}\phi_{J})},
\qquad \det W\neq0.
\end{equation*}
For coupled fields, non-singularity of individual diagonal blocks is not sufficient.
Complex fields and their conjugates are treated as independent variables when forming this Hessian.
This ensures that by means of the Legendre transformation, the Hamiltonian $\HC$ takes over the complete information on the given dynamical system from the Lagrangian $\LC$.
The definition of $\HC$ by Eq.~(\ref{H-def}) is referred to in literature as the ``De~Donder-Weyl'' Hamiltonian density.

Obviously, the dependencies of $\HC$ and $\LC$ on the $\phi_{I}$ and the $x^{\mu}$ only differ by a sign,
\begin{equation*}
\left.\pfrac{\HC}{x^{\mu}}\right\vert_{\text{expl}}=
-\left.\pfrac{\LC}{x^{\mu}}\right\vert_{\text{expl}},\qquad
\pfrac{\HC}{\phi_{I}}=-\pfrac{\LC}{\phi_{I}}=
-\pfrac{}{x^{\alpha}}\pfrac{\LC}{(\partial_{\alpha}\phi_{I})}=
-\pfrac{\pi_{I}^{\alpha}}{x^{\alpha}}.
\end{equation*}
These variables thus do not take part in the Legendre transformation of Eqs.~(\ref{p-def}), (\ref{H-def}).
Thus, with respect to this transformation, the Lagrangian density $\LC$ represents a function of the $\partial_{\mu}\phi_{I}$ only and does {\em not depend\/} on the canonical momenta $\pi_{I}^{\mu}$, whereas the Hamiltonian density $\HC$ is to be considered as a function of the $\pi_{I}^{\mu}$ only and does not depend on the derivatives $\partial_{\mu}\phi_{I}$ of the fields.
In order to derive the second canonical field equation, we calculate from Eq.~(\ref{H-def}) the partial derivative of $\HC$ with respect to $\pi_{I}^{\mu}$,
\begin{equation*}
\pfrac{\HC}{\pi_{I}^{\mu}}=\delta_{IJ}\,\delta_{\mu}^{\alpha}\,
\pfrac{\phi_{J}}{x^{\alpha}}=\pfrac{\phi_{I}}{x^{\mu}}
\qquad\Longleftrightarrow\qquad
\pfrac{\LC}{(\partial_{\mu}\phi_{I})}=\pi_{J}^{\alpha}
\delta_{JI}\,\delta_{\alpha}^{\mu}=\pi_{I}^{\mu}.
\end{equation*}
The complete set of covariant canonical field equations is thus given by
\begin{equation}\label{fgln}
\pfrac{\HC}{\pi_{I}^{\mu}}=\pfrac{\phi_{I}}{x^{\mu}},\qquad
\pfrac{\HC}{\phi_{I}}=-\pfrac{\pi_{I}^{\alpha}}{x^{\alpha}}.
\end{equation}
For a regular Legendre map, this pair of first-order partial differential equations is equivalent to the Euler-Lagrange equations~(\ref{elgl}), which are in general of second order.
We observe that in this formulation of the canonical field equations, all coordinates of space-time appear symmetrically --- similar to the Lagrangian formulation of Eq.~(\ref{elgl}).
Provided that the Lagrangian density $\LC$ is a Lorentz scalar, the dynamics of the fields is invariant with respect to Lorentz transformations.
The covariant Legendre transformation~(\ref{H-def}) passes this property to the Hamiltonian density $\HC$.
It thus ensures {\em a priori\/} the relativistic invariance of the fields that emerge as integrals of the canonical field equations if $\LC$ --- and hence $\HC$ --- represents a Lorentz scalar.
\section{\label{sec:d-w-cantra}Canonical transformations in covariant Hamiltonian field theory}
The covariant Legendre transformation~(\ref{H-def}) allows us to derive a canonical transformation theory in a way similar to that of point dynamics.
The main difference is that now the generating function of the canonical transformation is represented by a {\em vector\/} rather than by a scalar function.
The main benefit of this formalism is that we are not dealing with plain transformations.
Instead, we restrict ourselves {\em right from the beginning\/} to those transformations that preserve the form of the action functional.
This ensures the covariance of the canonical field equations.
A canonical transformation need not be a symmetry of a specified Hamiltonian.
That additional condition is discussed in Sect.~\ref{sec:gen-noether}.
Furthermore, with a generating function, we not only define the transformations of the fields but also pinpoint simultaneously the corresponding transformation law of the canonical momentum fields.
\subsection{Generating functions of type $\bF_{1}(\bphi,\bPhi,\bx)$}
Similar to the canonical formalism of point mechanics, we call a transformation of the fields $(\bphi,\bpi)\mapsto(\bPhi,\bPi)$ {\em canonical\/} if the form of the variational principle that is based on the action functional~(\ref{action-int}) is maintained,
\begin{equation}\label{varprinzip}
\delta\int_{R}\left(\pi_{I}^{\alpha}\pfrac{\phi_{I}}{x^{\alpha}}
-\HC(\bphi,\bpi,\bx)\right)d^{4}x\stackrel{!}{=}
\delta\int_{R}\left(\Pi_{I}^{\alpha}\pfrac{\Phi_{I}}{x^{\alpha}}
-\HC^{\prime}(\bPhi,\bPi,\bx)\right)d^{4}x.
\end{equation}
Equation~(\ref{varprinzip}) tells us that the {\em integrands\/} may differ by the divergence of a vector field $F_{1}^{\mu}$, whose variation vanishes on the boundary $\partial R$ of the integration region $R$ within space-time
\begin{equation*}
\delta\int_{R}\pfrac{F_{1}^{\alpha}}{x^{\alpha}}d^{4}x=
\delta\oint_{\partial R}F_{1}^{\alpha}dS_{\alpha}\stackrel{!}{=}0.
\end{equation*}
The immediate consequence of the form invariance of the variational principle is the form invariance of the covariant canonical field equations~(\ref{fgln})
\begin{equation*}
\pfrac{\HC^{\prime}}{\Pi_{I}^{\mu}}=\pfrac{\Phi_{I}}{x^{\mu}},\qquad
\pfrac{\HC^{\prime}}{\Phi_{I}}=-\pfrac{\Pi_{I}^{\alpha}}{x^{\alpha}}.
\end{equation*}
For the integrands of Eq.~(\ref{varprinzip}) --- hence for the Lagrangian densities $\LC$ and $\LC^{\prime}$ --- we thus obtain the condition
\begin{align}
\LC&=\LC^{\prime}+\pfrac{F_{1}^{\alpha}}{x^{\alpha}}\nonumber\\
\pi_{I}^{\alpha}\pfrac{\phi_{I}}{x^{\alpha}}-
\HC(\bphi,\bpi,\bx)&=\Pi_{I}^{\alpha}\pfrac{\Phi_{I}}{x^{\alpha}}
-\HC^{\prime}(\bPhi,\bPi,\bx)+\pfrac{F_{1}^{\alpha}}{x^{\alpha}}.
\label{intbed}
\end{align}
With the definition $F^{\mu}_{1}\equiv F^{\mu}_{1}(\bphi,\bPhi,\bx)$, we restrict ourselves to a function of exactly those arguments that now enter into transformation rules for the transition from the original to the new fields.
The divergence of $F^{\mu}_{1}$ writes, explicitly,
\begin{equation}\label{divF}
\pfrac{F_{1}^{\alpha}}{x^{\alpha}}=
\pfrac{F_{1}^{\alpha}}{\phi_{I}}\pfrac{\phi_{I}}{x^{\alpha}}+
\pfrac{F_{1}^{\alpha}}{\Phi_{I}}\pfrac{\Phi_{I}}{x^{\alpha}}+
{\left.\pfrac{F_{1}^{\alpha}}{x^{\alpha}}\right\vert}_{\text{expl}}.
\end{equation}
The rightmost term denotes the sum over the {\em explicit\/} dependence of the generating function $F^{\mu}_{1}$ on the $x^{\nu}$.
Comparing the coefficients of Eqs.~(\ref{intbed}) and (\ref{divF}), we find the local coordinate representation of the field transformation rules that are induced by the generating function $F^{\mu}_{1}$
\begin{equation}\label{genF1}
\pi_{I}^{\mu}=\pfrac{F_{1}^{\mu}}{\phi_{I}},\qquad
\Pi_{I}^{\mu}=-\pfrac{F_{1}^{\mu}}{\Phi_{I}},
\qquad \HC^{\prime}=\HC+{\left.\pfrac{F_{1}^{\alpha}}
{x^{\alpha}}\right\vert}_{\text{expl}}.
\end{equation}
The transformation rule for the Hamiltonian density implies that summation over $\alpha$ is to be performed.
In contrast to the transformation rule for the Lagrangian density $\LC$ of Eq.~(\ref{intbed}), the rule for the Hamiltonian density is determined by the {\em explicit\/} dependence of the generating function $F^{\mu}_{1}$ on the $x^{\nu}$.
Hence, if a generating function does not explicitly depend on the independent variables, $x^{\nu}$, then the {\em value\/} of the Hamiltonian density is not changed under the particular canonical transformation emerging thereof.

Differentiating the transformation rule for $\pi_{I}^{\mu}$ with respect to $\Phi_{J}$, and the rule for $\Pi_{J}^{\mu}$ with respect to $\phi_{I}$, we obtain a symmetry relation between original and transformed fields
\begin{equation*}
\pfrac{\pi_{I}^{\mu}}{\Phi_{J}}=
\pfrac{^{2}F_{1}^{\mu}}{\phi_{I}\partial\Phi_{J}}=
-\pfrac{\Pi_{J}^{\mu}}{\phi_{I}}.
\end{equation*}
The emerging of symmetry relations is a characteristic feature of {\em canonical\/} transformations.
As the symmetry relation directly follows from the second derivatives of the generating function, it does not apply for arbitrary transformations of the fields that do not follow from generating functions.
\subsection{\label{sec:genf2}Generating functions of type $\bF_{2}(\bphi,\bPi,\bx)$}
The generating function of a canonical transformation can alternatively be expressed in terms of a function of the original fields $\phi_{I}$ and of the new {\em conjugate\/} fields $\Pi_{I}^{\mu}$.
Where a common type-$F_{1}$ representation exists, the pertaining transformation rules can be related by the formal covariant Legendre transformation
\begin{equation}\label{legendre1}
F_{2}^{\mu}(\bphi,\bPi,\bx)=
F_{1}^{\mu}(\bphi,\bPhi,\bx)+\Phi_{J}\Pi_{J}^{\mu},
\qquad\Pi_{I}^{\mu}=-\pfrac{F_{1}^{\mu}}{\Phi_{I}}.
\end{equation}
By definition, the functions $F^{\mu}_{1}$ and $F^{\mu}_{2}$ agree with respect to their $\phi_{I}$ and $x^{\mu}$ dependencies
\begin{equation*}
\pfrac{F_{2}^{\mu}}{\phi_{I}}=\pfrac{F_{1}^{\mu}}{\phi_{I}}=\pi_{I}^{\mu},\qquad
\left.\pfrac{F_{2}^{\alpha}}{x^{\alpha}}\right\vert_{\text{expl}}=
\left.\pfrac{F_{1}^{\alpha}}{x^{\alpha}}\right\vert_{\text{expl}}=
\HC^{\prime}-\HC.
\end{equation*}
The variables $\phi_{I}$ and $x^{\mu}$ thus do not take part in the Legendre transformation from Eq.~(\ref{legendre1}).
Therefore, the two $F^{\mu}_{2}$-related transformation rules coincide with the respective rules derived previously from $F^{\mu}_{1}$.
As $F_{1}^{\mu}$ does not depend on the $\Pi_{I}^{\mu}$ whereas $F_{2}^{\mu}$ does not depend on the $\Phi_{I}$, the new transformation rule thus follows from the derivative of $F^{\mu}_{2}$ with respect to $\Pi_{J}^{\nu}$ as
\begin{equation*}
\pfrac{F_{2}^{\mu}}{\Pi_{I}^{\nu}}=
\Phi_{J}\pfrac{\Pi_{J}^{\mu}}{\Pi_{I}^{\nu}}=\Phi_{J}
\,\delta_{IJ}\,\delta_{\nu}^{\mu}.
\end{equation*}
We thus end up with set of transformation rules
\begin{equation}\label{genF2}
\pi_{I}^{\mu}=\pfrac{F_{2}^{\mu}}{\phi_{I}},\qquad
\Phi_{I}\,\delta_{\nu}^{\mu}=\pfrac{F_{2}^{\mu}}{\Pi_{I}^{\nu}},
\qquad\HC^{\prime}=\HC+{\left.\pfrac{F_{2}^{\alpha}}
{x^{\alpha}}\right\vert}_{\text{expl}},
\end{equation}
which agrees with the set~(\ref{genF1}) wherever the relations in Eq.~(\ref{legendre1}) admit a common local solution for all space-time components.
Non-singularity of individual mixed-derivative matrices of $F_{1}^{\mu}$ does not by itself establish this compatibility.
In particular, an admissible type-$F_{2}$ generator need not admit a non-singular type-$F_{1}$ representation.
The rules~(\ref{genF2}) can also be obtained directly by comparing the canonical action integrands.
From the second partial derivations of $F^{\mu}_{2}$ one immediately derives the symmetry relation
\begin{equation*}
\pfrac{\pi_{I}^{\mu}}{\Pi_{J}^{\nu}}=
\pfrac{^{2}F_{2}^{\mu}}{\phi_{I}\partial\Pi_{J}^{\nu}}=
\pfrac{\Phi_{J}}{\phi_{I}}\,\delta_{\nu}^{\mu},
\end{equation*}
which is a necessary consequence of the generating-function rules.

The function $F_{2}^{\mu}$ is not arbitrary.
We call it {\em admissible\/} if all the component equations~(\ref{genF2}) define one common, locally invertible transformation on the stated phase-space domain.
For a twice continuously differentiable generator on an open domain with independent polymomenta, the off-diagonal conditions imply
\begin{equation*}
\pfrac{F_{2}^{\mu}}{\Pi_{I}^{\nu}}=0\quad(\nu\ne\mu).
\end{equation*}
Thus, $F_{2}^{\mu}$ depends only on the momentum column $\Pi_{I}^{\mu}$.
Its diagonal derivative must nevertheless yield the same $\Phi_{I}$ for every $\mu$.
In more than one space-time dimension, differentiating these conditions with respect to the other momentum columns shows that $\Phi_{I}$ is independent of all polymomenta.
Consequently, the admissible local form is
\begin{equation*}
F_{2}^{\mu}=\Pi_{I}^{\mu}f_{I}(\bphi,\bx)+b^{\mu}(\bphi,\bx),
\qquad \Phi_{I}=f_{I}(\bphi,\bx),
\qquad \det\left(\pfrac{f_{I}}{\phi_{J}}\right)\ne0.
\end{equation*}
The corresponding momentum transformation is
\begin{equation*}
\pi_{I}^{\mu}=\Pi_{J}^{\mu}\pfrac{f_{J}}{\phi_{I}}
+\pfrac{b^{\mu}}{\phi_{I}}.
\end{equation*}
Hence this fixed-space-time class describes field-space point transformations and affine transformations of the momenta.
For constrained momentum tensors, the differentiation rules must instead be restricted to the independent components, as illustrated in the Maxwell example below.
\section{\label{sec:examples-ham} Examples for Hamiltonian densities in covariant field theory}
We present some simple examples of Hamiltonian densities as they emerge from Lagrangian densities of classical Lagrangian field theory.
It is shown that resulting canonical field equations are equivalent to the corresponding Euler-Lagrange equations.
\subsection{Klein-Gordon Hamiltonian density for complex fields}
We first consider the Klein-Gordon {\em Lagrangian density\/} $\LC_{\text{KG}}$ for a {\em complex\/} scalar field $\phi$ that is associated with mass $m$ (see, for instance, Ref.~\cite{greiner}), given by
\begin{equation*}
\LC_{\text{KG}}\left(\phi,\phi^{*},\partial^{\mu}\phi,
\partial_{\mu}\phi^{*}\,\right)=\pfrac{\phi^{*}}{x^{\alpha}}
\pfrac{\phi}{x_{\alpha}}-m^{2}\,\phi^{*}\phi.
\end{equation*}
Herein, $\phi^{*}$ denotes complex conjugate field of $\phi$.
Both quantities are to be treated as independent.
With $[L]$ denoting the dimension of ``length,'' we have with $\hbar=c=1$, i.e.\ in ``natural units'', $[\LC]=[L]^{-4}$, $[m]=[L]^{-1}$, and $[\partial_{\mu}]=[L]^{-1}$ so that $[\phi]=[L]^{-1}$.
The Euler-Lagrange equations~(\ref{elgl}) for $\phi$ and $\phi^{*}$ follow from this Lagrangian density as
\begin{equation}\label{ex3-fg1}
\pfrac{^{2}}{x_{\alpha}\partial x^{\alpha}}\phi^{*}=
-m^{2}\,\phi^{*},\qquad
\pfrac{^{2}}{x_{\alpha}\partial x^{\alpha}}\phi=-m^{2}\,\phi.
\end{equation}
As a prerequisite for deriving the corresponding Hamiltonian density $\HC_{\text{KG}}$ we must first define from $\LC_{\text{KG}}$ the conjugate momentum fields,
\begin{equation*}
\pi^{\mu}=\pfrac{\LC_{\text{KG}}}{\left(\partial_{\mu}\phi^{*}\,\right)}=
\pfrac{\phi}{x_{\mu}},\qquad
\pi_{\mu}^{*}=\pfrac{\LC_{\text{KG}}}{(\partial^{\mu}\phi)}=
\pfrac{\phi^{*}}{x^{\mu}},
\end{equation*}
which means that $[\pi^{\mu}]=[L]^{-2}$.
The full Hessian for the independent fields $\phi,\phi^{*}$ is non-singular.
Its diagonal blocks vanish and its mixed blocks are invertible.
In the index convention used here,
\begin{equation*}
\det\left(\pfrac{^{2}\LC_{\text{KG}}}{(\partial^{\mu}\phi)
\partial\left(\partial_{\nu}\phi^{*}\,\right)}\right)=
\det\left(\pfrac{\pi_{\mu}^{*}}{\left(
\partial_{\nu}\phi^{*}\,\right)}\right)=
\det\left(\delta_{\mu}^{\nu}\right)=1.
\end{equation*}
For a Lagrangian density that is {\em quadratic\/} in the field derivatives, regularity requires the full quadratic coefficient matrix to have full rank.
Quadratic dependence alone is insufficient.
The Hamiltonian density $\HC$ then follows as the Legendre transform of the Lagrangian density
\begin{equation*}
\HC(\pi^{\mu},\pi_{\mu}^{*},\phi,\phi^{*})=
\pi_{\alpha}^{*}\pfrac{\phi}{x_{\alpha}}+
\pfrac{\phi^{*}}{x^{\alpha}}\pi^{\alpha}-
\LC(\partial^{\mu}\phi,\partial_{\mu}\phi^{*},\phi,\phi^{*}),
\end{equation*}
thus $[\HC]=[\LC]=[L]^{-4}$.
The Klein-Gordon {\em Hamiltonian density\/} $\HC_{\text{KG}}$ is then given by
\begin{equation}\label{ex3-hd}
\HC_{\text{KG}}(\pi_{\mu},\pi_{\mu}^{*},\phi,\phi^{*})=
\pi_{\alpha}^{*}\pi^{\alpha}+m^{2}\,\phi^{*}\phi.
\end{equation}
For the Hamiltonian density~(\ref{ex3-hd}), the canonical field equations~(\ref{fgln}) provide the following set of coupled first order partial differential equations
\begin{align*}
\pfrac{\phi^{*}}{x^{\mu}}=\pfrac{\HC_{\text{KG}}}{\pi^{\mu}}&=\pi_{\mu}^{*},&
\pfrac{\phi}{x_{\mu}}=\pfrac{\HC_{\text{KG}}}{\pi_{\mu}^{*}}&=\pi^{\mu}\\
-\pfrac{\pi_{\alpha}^{*}}{x_{\alpha}}=\pfrac{\HC_{\text{KG}}}{\phi}&=m^{2}\phi^{*},&
-\pfrac{\pi^{\alpha}}{x^{\alpha}}=\pfrac{\HC_{\text{KG}}}{\phi^{*}}&=m^{2}\phi.
\end{align*}
In the first row, the canonical field equations for the scalar fields $\phi$ and $\phi^{*}$ reproduce the definitions of the momentum fields $\pi^{\mu}$ and $\pi_{\mu}^{*}$ from the Lagrangian density $\LC_{\text{KG}}$.
Eliminating the $\pi^{\mu}$, $\pi_{\mu}^{*}$ from the canonical field equations then yields the Euler-Lagrange equations of Eq.~(\ref{ex3-fg1}).
\subsection{\label{sec:maxwell}Maxwell's equations as canonical field equations}
The Lagrangian density $\LC_{\text{M}}$ of the electromagnetic field is given by
\begin{equation}\label{ld-maxwell}
\LC_{\text{M}}(\ba,\bpartial\ba,\bx)=-\quarter f_{\alpha\beta}f^{\alpha\beta}-
j^{\alpha}(\bx)\,a_{\alpha},\quad
f_{\mu\nu}=\pfrac{a_{\nu}}{x^{\mu}}-\pfrac{a_{\mu}}{x^{\nu}}.
\end{equation}
Herein, the four components $a^{\mu}$ of the $4$-vector potential $\ba$ now take the place of the scalar fields $\phi_{I}\equiv a^{\mu}$ in the notation used so far.
The Lagrangian density (\ref{ld-maxwell}) thus entails a set of {\em four\/} Euler-Lagrange equations, i.e., an equation for each component $a_{\mu}$.
The source vector $\bj=(\rho,j_{x},j_{y},j_{z})$ denotes the $4$-vector of electric currents combining the usual current density vector $(j_{x},j_{y},j_{z})$ of configuration space with the charge density $\rho$.
In this example, $j^{\mu}(\bx)$ is a prescribed external current.
It is assumed to be conserved, $\partial_{\mu}j^{\mu}=0$, as also required by the Maxwell equations.
In a local Lorentz frame, i.e., in Minkowski space, the Euler-Lagrange equations~(\ref{elgl}) take on the form,
\begin{equation}\label{elgl1}
\pfrac{}{x^{\alpha}}\pfrac{\LC_{\text{M}}}{(\partial_{\alpha}a_{\mu})}-
\pfrac{\LC_{\text{M}}}{a_{\mu}}=0,\qquad\mu=0,\ldots,3.
\end{equation}
With $\LC_{\text{M}}$ from Eq.~(\ref{ld-maxwell}), we obtain directly
\begin{equation}\label{el-maxwell}
\pfrac{f^{\mu\alpha}}{x^{\alpha}}+j^{\mu}=0.
\end{equation}
In Minkowski space, this is the tensor form of the inhomogeneous Maxwell equation.
In order to formulate the equivalent Hamiltonian description, we first define, according to Eq.~(\ref{p-def}), the canonical momentum field components $p^{\mu\nu}$ as the conjugate objects of the derivatives of the $4$-vector potential $\ba$
\begin{equation}\label{p-def1}
p^{\mu\nu}=\pfrac{\LC_{\text{M}}}{(\partial_{\nu}a_{\mu})}
\equiv\pfrac{\LC_{\text{M}}}{a_{\mu,\nu}}
\end{equation}
With the particular Lagrangian density~(\ref{ld-maxwell}), Eq.~(\ref{p-def1}) means
\begin{align*}
f_{\alpha\beta}&=a_{\beta,\alpha}-a_{\alpha,\beta}\\
p^{\mu\nu}&=-\quarter\left(
\pfrac{f_{\alpha\beta}}{a_{\mu,\nu}}f^{\alpha\beta}+
\pfrac{f^{\alpha\beta}}{a_{\mu,\nu}}f_{\alpha\beta}
\right)=-\onehalf\pfrac{f_{\alpha\beta}}{a_{\mu,\nu}}f^{\alpha\beta}\\
&=-\onehalf\left(\delta_{\beta}^{\mu}\delta_{\alpha}^{\nu}-
\delta_{\alpha}^{\mu}\delta_{\beta}^{\nu}\right)f^{\alpha\beta}
=\onehalf(f^{\mu\nu}-f^{\nu\mu})\\
&=f^{\mu\nu}.
\end{align*}
The tensor $p^{\mu\nu}$ thus matches exactly the electromagnetic field tensor $f^{\mu\nu}$ from Eq.~(\ref{ld-maxwell}) and hence inherits the skew-symmetry of $f^{\mu\nu}$ because of the particular dependence of $\LC_{\mathrm{M}}$ on the $a_{\mu,\nu}\equiv\partial a_{\mu}/\partial x^{\nu}$.
The full derivative Hessian is singular in the symmetric derivative sector.
Accordingly, the following Hamiltonian description is restricted to $p^{(\mu\nu)}=0$.
It is not a regular Legendre transformation on all sixteen components of $a_{\mu,\nu}$.
Momentum derivatives are normalized by $\delta\HC=(\partial\HC/\partial p^{\mu\nu})\delta p^{\mu\nu}$ with antisymmetric variations and summation over all ordered pairs.

As the Lagrangian density~(\ref{ld-maxwell}) now describes the dynamics of a {\em vector field}, $a_{\mu}$, rather than a set of scalar fields $\phi_{I}$, the canonical momenta $p^{\mu\nu}$ now constitute a second rank {\em tensor\/} rather than a vector.
The Legendre transformation corresponding to Eq.~(\ref{H-def}) then comprises the product $p^{\alpha\beta}\partial_{\beta}a_{\alpha}$.
The skew-symmetry of the momentum tensor $p^{\mu\nu}$ picks out the skew-symmetric part of $\partial_{\nu}a_{\mu}$ as the symmetric part of $\partial_{\nu}a_{\mu}$ vanishes identically calculating the product $p^{\alpha\beta}\partial_{\beta}a_{\alpha}$
\begin{equation*}
p^{\alpha\beta}\pfrac{a_{\alpha}}{x^{\beta}}=
\onehalf p^{\alpha\beta}\underbrace{\left(\pfrac{a_{\alpha}}{x^{\beta}}-
\pfrac{a_{\beta}}{x^{\alpha}}\right)}_{=f_{\beta\alpha}}+\onehalf
\underbrace{p^{\alpha\beta}\left(\pfrac{a_{\alpha}}{x^{\beta}}+
\pfrac{a_{\beta}}{x^{\alpha}}\right)}_{\equiv0}.
\end{equation*}
For a skew-symmetric momentum tensor $p^{\mu\nu}$, we thus obtain the Hamiltonian density $\HC_{\text{M}}$ as the Legendre-transformed Lagrangian density $\LC_{\text{M}}$
\begin{equation*}
\HC_{\text{M}}(\ba,\bp,\bx)=-\onehalf p^{\alpha\beta}f_{\alpha\beta}-
\LC_{\text{M}}(\ba,\bpartial\ba,\bx).
\end{equation*}
From this restricted Legendre transformation prescription and the corresponding Euler-La\-grange equations~(\ref{elgl1}), the canonical field equations are immediately obtained as
\begin{align*}
\pfrac{\HC_{\text{M}}}{p^{\mu\nu}}&=-\frac{1}{2}f_{\mu\nu}=\frac{1}{2}\left(
\pfrac{a_{\mu}}{x^{\nu}}-\pfrac{a_{\nu}}{x^{\mu}}\right)\\
\pfrac{\HC_{\text{M}}}{a_{\mu}}&=-\pfrac{\LC_{\text{M}}}{a_{\mu}}=
-\pfrac{}{x^{\alpha}}\pfrac{\LC_{\text{M}}}{(\partial_{\alpha}a_{\mu})}=
-\pfrac{p^{\mu\alpha}}{x^{\alpha}}\\
\pfrac{\HC_{\text{M}}}{x^{\nu}}&=-\pfrac{\LC_{\text{M}}}{x^{\nu}}.
\end{align*}
The Hamiltonian density for the Lagrangian density~(\ref{ld-maxwell}) follows as
\begin{align}
\HC_{\text{M}}(\ba,\bp,\bx)&=-\onehalf p^{\alpha\beta}p_{\alpha\beta}+
\quarter p^{\alpha\beta}p_{\alpha\beta}+j^{\alpha}(\bx)\,
a_{\alpha}\nonumber\\
&=-\quarter p^{\alpha\beta}p_{\alpha\beta}+
j^{\alpha}(\bx)\,a_{\alpha}.
\label{hd-maxwell}
\end{align}
The first canonical field equation follows from the derivative of the Hamiltonian density~(\ref{hd-maxwell}) with respect to $p^{\mu\nu}$ and $p_{\mu\nu}$
\begin{equation}\label{fg1-maxwell}
\frac{1}{2}\left(\pfrac{a_{\mu}}{x^{\nu}}-\pfrac{a_{\nu}}{x^{\mu}}\right)=
\pfrac{\HC_{\text{M}}}{p^{\mu\nu}}=-\onehalf p_{\mu\nu},\quad
\frac{1}{2}\left(\pfrac{a^{\mu}}{x_{\nu}}-\pfrac{a^{\nu}}{x_{\mu}}\right)=
\pfrac{\HC_{\text{M}}}{p_{\mu\nu}}=-\onehalf p^{\mu\nu},
\end{equation}
which reproduces the definition of $p_{\mu\nu}$ and $p^{\mu\nu}$ from Eq.~(\ref{p-def1}).

The second canonical field equation is obtained calculating the derivative of the Hamiltonian density~(\ref{hd-maxwell}) with respect to $a_{\mu}$
\begin{equation*}
-\pfrac{p^{\mu\alpha}}{x^{\alpha}}=
\pfrac{\HC_{\text{M}}}{a_{\mu}}=j^{\mu}.
\end{equation*}
Inserting the first canonical equation, the second order field equation for the $a_{\mu}$ is thus obtained for the Maxwell Hamiltonian density~(\ref{hd-maxwell}) as
\begin{equation*}
\pfrac{f^{\mu\alpha}}{x^{\alpha}}+j^{\mu}=0,
\end{equation*}
which agrees, as expected, with the corresponding Euler-Lagrange equation~(\ref{el-maxwell}).
\subsection{The Proca Hamiltonian density}
In relativistic quantum field theory, the dynamics of particles of spin $1$ and mass $m$ is derived from the Proca Lagrangian density $\LC_{\text{P}}$,
\begin{equation*}
\LC_{\text{P}}=-\quarter f^{\alpha\beta}f_{\alpha\beta}+
\onehalf m^{2}a^{\alpha}a_{\alpha},\qquad
f_{\mu\nu}=\pfrac{a_{\nu}}{x^{\mu}}-\pfrac{a_{\mu}}{x^{\nu}}.
\end{equation*}
We observe that the kinetic term of $\LC_{\text{P}}$ agrees with that of the Lagrangian density $\LC_{\text{M}}$ of the electromagnetic field of Eq.~(\ref{ld-maxwell}).
Therefore, the field equations emerging from the Euler-Lagrange equations~(\ref{elgl1}) are similar to those of Eq.~(\ref{el-maxwell})
\begin{equation}\label{el-proca}
\pfrac{f^{\mu\alpha}}{x^{\alpha}}-m^{2}a^{\mu}=0.
\end{equation}
Thus $[\LC]=[L]^{-4}$, $[m]=[L]^{-1}$, and $[\partial_{\mu}]=[L]^{-1}$ entails a dimension of the $4$-vector fields $[\ba]=[L]^{-1}$ and $[\pmb{f}]=[L]^{-2}$ in natural units.
The transition to the corresponding Hamilton description is performed by defining on the basis of the actual Lagrangian $\LC_{\text{P}}$ the canonical momentum field tensors $p^{\mu\nu}$ as the conjugate objects of the derivatives of the $4$-vector potential $\ba$
\begin{equation*}
p^{\mu\nu}=\pfrac{\LC_{\text{P}}}{\left(\partial_{\nu}a_{\mu}\right)}
\equiv\pfrac{\LC_{\text{P}}}{a_{\mu,\nu}}.
\end{equation*}
Similar to the preceding section, we find
\begin{equation*}
p^{\mu\nu}=f^{\mu\nu},\qquad p_{\mu\nu}=f_{\mu\nu},\qquad [\pmb{p}]=[\pmb{f}]=[L]^{-2},
\end{equation*}
because of the particular dependence of $\LC_{\text{P}}$ on the derivatives of the $a^{\mu}$.
With $p^{\alpha\beta}$ being skew-symmetric in $\alpha,\beta$, the product $p^{\alpha\beta}\,a_{\alpha,\beta}$ picks out the skew-symmetric part of the partial derivative $\partial a_{\alpha}/\partial x^{\beta}$ as the product with the symmetric part vanishes identically.
Denoting the skew-symmetric part by $a_{[\alpha,\beta]}$, the Legendre transformation prescription
\begin{align*}
\HC_{\text{P}}&=p^{\alpha\beta}\,a_{\alpha,\beta}-\LC_{\text{P}}=
p^{\alpha\beta}\,a_{[\alpha,\beta]}-\LC_{\text{P}}\\
&=\onehalf p^{\alpha\beta}\left(\pfrac{a_{\alpha}}{x^{\beta}}-
\pfrac{a_{\beta}}{x^{\alpha}}\right)-\LC_{\text{P}},
\end{align*}
leads to the Proca Hamiltonian density by following the path of Eq.~(\ref{hd-maxwell})
\begin{equation}\label{hd-proca}
\HC_{\text{P}}=-\quarter p^{\alpha\beta}p_{\alpha\beta}-
\onehalf m^{2}a^{\alpha}a_{\alpha}.
\end{equation}
The canonical field equations emerge as
\begin{align*}
a_{[\mu,\nu]}\equiv\frac{1}{2}\left(\pfrac{a_{\mu}}{x^{\nu}}-
\pfrac{a_{\nu}}{x^{\mu}}\right)=
\pfrac{\HC_{\text{P}}}{p^{\mu\nu}}&=-\onehalf p_{\mu\nu}\\
-\pfrac{p^{\mu\alpha}}{x^{\alpha}}=\pfrac{\HC_{\text{P}}}{a_{\mu}}&=
-m^{2}a^{\mu}.
\end{align*}
By means of eliminating $p^{\mu\nu}$, this coupled set of first order equations can be converted into second order equations for the vector field $\ba(\bx)$,
\begin{equation*}
\pfrac{}{x_{\alpha}}\left(\pfrac{a_{\mu}}{x^{\alpha}}-
\pfrac{a_{\alpha}}{x^{\mu}}\right)+m^{2}a_{\mu}=0,
\end{equation*}
which coincides with the Euler-Lagrange equation~(\ref{el-proca}).
For $m\ne0$, its divergence implies $\partial_{\mu}a^{\mu}=0$, and the remaining equation is \mbox{$(\Box+m^{2})a_{\mu}=0$}.
The Proca mass term does not remove the degeneracy of the derivative Hessian, so the restriction to antisymmetric momenta remains necessary.
\subsection{\label{sec:dirac-ham}The Dirac Hamiltonian density}
The dynamics of particles with spin $\frac{1}{2}$ and mass $m$ is described by the Dirac equation.
With $\gamma^{\mu}$, $\mu=0,\ldots,3$ denoting the $4\times 4$ Dirac matrices, and $\psi$ a four component Dirac spinor, the Dirac Lagrangian density $\LC_{\text{D}}$ is given by
\begin{equation}\label{ld-dirac}
\LC_{\text{D}}=i\bar{\psi}\gamma^{\alpha}
\pfrac{\psi}{x^{\alpha}} - m\,\bar{\psi}\psi,
\end{equation}
wherein $\bar{\psi}\equiv\psi^{\dagger}\gamma^{0}$ denotes the adjoint spinor of $\psi$.
In the following we summarize some fundamental relations that apply for the Dirac matrices $\gamma^{\mu}$, and their duals, $\gamma_{\mu}$,
\begin{align}
\{\gamma^{\mu},\gamma^{\nu}\}&\equiv\gamma^{\mu}\gamma^{\nu}+
\gamma^{\nu}\gamma^{\mu}=2\eta^{\mu\nu}\Eins\nonumber\\
\gamma^{\alpha}\gamma_{\alpha}&=\gamma_{\alpha}\gamma^{\alpha}=4\;\Eins\nonumber\\
\left[\gamma^{\mu},\gamma^{\nu}\right]&\equiv
\gamma^{\mu}\gamma^{\nu}-\gamma^{\nu}\gamma^{\mu}\equiv
-2i\,\sigma^{\mu\nu}\nonumber\\
\left[\gamma_{\mu},\gamma_{\nu}\right]&\equiv
\gamma_{\mu}\gamma_{\nu}\,\,-\gamma_{\nu}\gamma_{\mu}\,\equiv
-2i\,\sigma_{\mu\nu}\nonumber\\
\det\sigma^{\mu\nu}&=1,\qquad\mu\ne\nu\nonumber\\
\tau_{\mu\alpha}\sigma^{\alpha\nu}&=\sigma^{\nu\alpha}\tau_{\alpha\mu}=
\delta_{\mu}^{\nu}\,\Eins\nonumber\\
\gamma^{\alpha}\tau_{\alpha\mu}&=\tau_{\mu\alpha}\gamma^{\alpha}=
-\frac{i}{3}\,\gamma_{\mu}\nonumber\\
\gamma_{\alpha}\sigma^{\alpha\mu}&=\sigma^{\mu\alpha}\gamma_{\alpha}=
3i\,\gamma^{\mu}\nonumber\\
\gamma^{\alpha}\tau_{\alpha\beta}\gamma^{\beta}&=-\frac{4i}{3}\Eins\nonumber\\
\gamma_{\alpha}\sigma^{\alpha\beta}\gamma_{\beta}&=12i\,\Eins,\qquad
\sigma^{\alpha\beta}\,\sigma_{\alpha\beta}=12\,\Eins\nonumber\\
3\tau_{\mu\nu}+\sigma_{\mu\nu}&=2i\,\eta_{\mu\nu}\,\Eins.
\label{dirac-algebra}
\end{align}
Herein, the symbol $\Eins$ stands for the $4\times 4$ unit matrix, and the real numbers $\eta^{\mu\nu},\eta_{\mu\nu}\in\RB$ for an element of the Minkowski metric $(\eta^{\mu\nu})=(\eta_{\mu\nu})$.
The matrices $(\sigma^{\mu\nu})$ and $(\tau_{\mu\nu})$ are to be understood as $4\times 4$ block matrices, with each block $\sigma^{\mu\nu}$, $\tau_{\mu\nu}$ representing a $4\times 4$ matrix of complex numbers.
Thus, $(\sigma^{\mu\nu})$ and $(\tau_{\mu\nu})$ are actually $16\times 16$ matrices of complex numbers.

Natural units are defined by setting $\hbar=c=1$.
Denoting ``the dimension of'' by the symbol ``$[]$'', we then have for the dimension of the mass $m$, length $L$, time $T$, and energy $E$
\begin{equation*}
[m]=[L]^{-1}=[T]^{-1}=[E].
\end{equation*}
Then
\begin{equation*}
[\LC_{\text{D}}]=[L]^{-4},\qquad [\psi]=[L]^{-3/2},\qquad [\partial_{\mu}]=[m]=[L]^{-1}.
\end{equation*}
The Dirac Lagrangian density $\LC_{\text{D}}$ can be rendered symmetric by combining the Lagrangian density Eq.~(\ref{ld-dirac}) with its adjoint, which leads to
\begin{equation}\label{ld-dirac-symm}
\LC_{\text{D}}=\frac{i}{2}\left(\bar{\psi}\gamma^{\alpha}
\pfrac{\psi}{x^{\alpha}}-\pfrac{\bar{\psi}}{x^{\alpha}}\gamma^{\alpha}\psi\right)
-m\bar{\psi}\psi.
\end{equation}
The resulting Euler-Lagrange equations are identical to those derived from Eq.~(\ref{ld-dirac}),
\begin{align}
i\gamma^{\alpha}\pfrac{\psi}{x^{\alpha}}-m\psi&=0\nonumber\\
i\pfrac{\bar{\psi}}{x^{\alpha}}\gamma^{\alpha}+
m\bar{\psi}&=0.
\label{el-dirac}
\end{align}
As both Lagrangians~(\ref{ld-dirac}) and~(\ref{ld-dirac-symm}) are {\em linear\/} in the derivatives of the fields, the determinant of the Hessian vanishes,
\begin{equation}\label{ld-irregular}
\det\left[\pfrac{^{2}\LC_{\text{D}}}
{\left(\partial_{\mu}\psi\right)\partial
\left(\partial_{\nu}\bar{\psi}\right)}\right]=0.
\end{equation}
Therefore, Legendre transformations of the Lagrangian densities~(\ref{ld-dirac}) and~(\ref{ld-dirac-symm}) are irregular.
Nevertheless, as a Lagrangian density is determined only up to the divergence of an arbitrary vector function $F^{\mu}$ according to Eq.~(\ref{intbed}), one can construct an equivalent Lagrangian density $\LC_{\text{D}}^{\prime}$ that yields identical Euler-Lagrange equations while yielding a regular Legendre transformation.
Here, equivalence refers to the bulk equations with compactly supported variations, or boundary conditions for which the variation of the added boundary integral vanishes.
Since the vector below depends on field derivatives, this change of Lagrangian representative is not itself a type-$F_{1}$ or type-$F_{2}$ generator of Sect.~\ref{sec:d-w-cantra}.
The additional term\cite{gasi} emerges as the divergence of a vector function $F^{\mu}$, which may be expressed in symmetric form as
\begin{equation*}
F^{\mu}=\frac{i}{6m^{*}}\left(\bar{\psi}\,\sigma^{\mu\alpha}\pfrac{\psi}{x^{\alpha}}+
\pfrac{\bar{\psi}}{x^{\alpha}}\sigma^{\alpha\mu}\,\psi\right),\qquad
[\pmb{F}]=[L]^{-3}.
\end{equation*}
The regularization mass scale $m^{*}\in\RB\setminus\{0\}$ is a real, nonzero constant with the natural dimension of mass.
It is not a gauge-fixing parameter.
The free field equations are independent of its value, whereas its role after introducing gauge interactions is discussed in Sect.~\ref{sec:pauli-comparison}.
Explicitly, the additional term is given by
\begin{align*}
\pfrac{F^{\beta}}{x^{\beta}}&=\frac{i}{6m^{*}}\left(
\partial_{\beta}\bar{\psi}\sigma^{\beta\alpha}\partial_{\alpha}\psi+
\bar{\psi}\sigma^{\beta\alpha}\partial_{\beta}\partial_{\alpha}\psi+
\partial_{\beta}\partial_{\alpha}\bar{\psi}\sigma^{\alpha\beta}\psi+
\partial_{\alpha}\bar{\psi}\sigma^{\alpha\beta}\partial_{\beta}\psi\right)\\
&=\pfrac{\bar{\psi}}{x^{\alpha}}\,\frac{i\sigma^{\alpha\beta}}{3m^{*}}
\pfrac{\psi}{x^{\beta}}.
\end{align*}
Note that the double sums $\sigma^{\beta\alpha}\partial_{\beta}\partial_{\alpha}\psi$ and $\partial_{\beta}\partial_{\alpha}\bar{\psi}\sigma^{\alpha\beta}$ vanish identically, as we sum over a symmetric ($\partial_{\mu}\partial_{\nu}\psi=\partial_{\nu}\partial_{\mu}\psi$) and a skew-symmetric ($\sigma^{\mu\nu}=-\sigma^{\nu\mu}$) factor.
Following Eq.~(\ref{intbed}), the equivalent Lagrangian density is given by $\LC_{\text{D}}^{\prime}=\LC_{\text{D}}+\partial F^{\beta}/\partial x^{\beta}$, which means, explicitly,
\begin{equation}\label{ld-dirac-regular}
\LC_{\text{D}}^{\prime}=\frac{i}{2}\left(\bar{\psi}\gamma^{\alpha}
\pfrac{\psi}{x^{\alpha}}-\pfrac{\bar{\psi}}{x^{\alpha}}\gamma^{\alpha}\psi\right)
+\pfrac{\bar{\psi}}{x^{\alpha}}\,\frac{i\sigma^{\alpha\beta}}{3m^{*}}
\pfrac{\psi}{x^{\beta}}-m\,\bar{\psi}\psi.
\end{equation}
Due to the skew-symmetry of the $\sigma^{\mu\nu}$, the Euler-Lagrange equations~(\ref{elgl}) for $\LC_{\text{D}}^{\prime}$ yield again the Dirac equations~(\ref{el-dirac}).
We remark that the regularized Dirac Lagrangian~(\ref{ld-dirac-regular}) can equivalently be written as
\begin{equation*}
\LC_{\text{D}}^{\prime}=\left(\pfrac{\bar{\psi}}{x^{\alpha}}-
\frac{im^{*}}{2}\bar{\psi}\gamma_{\alpha}\right)\frac{i\sigma^{\alpha\beta}}{3m^{*}}
\left(\pfrac{\psi}{x^{\beta}}+\frac{im^{*}}{2}\gamma_{\beta}\psi\right)+
\left(m^{*}-m\right)\bar{\psi}\psi.
\end{equation*}
This representation of the Dirac Lagrangian will be recognized as the analogue of the Dirac Hamiltonian $\HC_{\text{D}}$ to be derived in Eq.~(\ref{hd-dirac}).

As desired, the full derivative Hessian of $\LC_{\text{D}}^{\prime}$ is non-singular.
Its mixed $16\times16$ block, including the spinor indices, satisfies
\begin{equation}\label{ld-regular}
\begin{split}
\pfrac{^{2}\LC_{\text{D}}^{\prime}}
{\left(\partial_{\mu}\bar{\psi}\right)
\partial\left(\partial_{\nu}\psi\right)}
&=\frac{i\sigma^{\mu\nu}}{3m^{*}},
\qquad \det\left(\frac{i\sigma^{\mu\nu}}{3m^{*}}\right)\ne0,\\
\frac{3m^{*}}{i}\tau_{\lambda\mu}
\frac{i\sigma^{\mu\nu}}{3m^{*}}
&=\delta_{\lambda}^{\nu}\,\Eins.
\end{split}
\end{equation}
The inverse follows from Eq.~(\ref{dirac-algebra}), not from invertibility of the individual $4\times4$ blocks $\sigma^{\mu\nu}$.
With $\psi$ and $\bar\psi$ treated as independent fields, the full Hessian has this invertible mixed block and its transpose as its off-diagonal blocks.
Thus, the Legendre transformation of the Lagrangian density $\LC_{\text{D}}^{\prime}$ is now {\em regular}.
It is remarkable that it is exactly a term which does {\em not\/} contribute to the Euler-Lagrange equations that makes the Legendre transformation of $\LC_{\text{D}}^{\prime}$ {\em regular\/} and thus transfers the information on the dynamical system that is contained in the Lagrangian to the Hamiltonian description.
The canonical momenta follow as
\begin{align}
\bar{\pi}^{\mu}&=
\pfrac{\LC_{\text{D}}^{\prime}}{\left(\partial_{\mu}
\psi\right)}=\hphantom{-}\frac{i}{2}\bar{\psi}\gamma^{\mu}+
\pfrac{\bar{\psi}}{x^{\alpha}}\,\frac{i\sigma^{\alpha\mu}}{3m^{*}}
\nonumber\\
\pi^{\mu}&=
\pfrac{\LC_{\text{D}}^{\prime}}{\left(\partial_{\mu}
\bar{\psi}\right)}=-\frac{i}{2}\gamma^{\mu}\psi+
\frac{i\sigma^{\mu\alpha}}{3m^{*}}\pfrac{\psi}{x^{\alpha}},
\label{pi-dirac}
\end{align}
which states that $[\pi^{\mu}]=[\psi]=[L]^{-3/2}$.
The Legendre transformation can now be worked out, yielding
\begin{align*}
\HC_{\text{D}}&=\bar{\pi}^{\alpha}\pfrac{\psi}{x^{\alpha}}+
\pfrac{\bar{\psi}}{x^{\alpha}}\,\pi^{\alpha}-
\LC_{\text{D}}^{\prime}\\
&=\pfrac{\bar{\psi}}{x^{\alpha}}\,\frac{i\sigma^{\alpha\beta}}{3m^{*}}
\pfrac{\psi}{x^{\beta}}+m\,\bar{\psi}\,\psi\\
&=\left(\bar{\pi}^{\beta}-\frac{i}{2}\bar{\psi}\gamma^{\beta}
\right)\pfrac{\psi}{x^{\beta}}+m\,\bar{\psi}\,\psi,
\end{align*}
thus $[\HC_{\text{D}}]=[\LC_{\text{D}}]=[L]^{-4}$.
As the Hamiltonian density must always be expressed in terms of the canonical momenta rather than by the field derivatives, we must solve Eq.~(\ref{pi-dirac}) for $\partial_{\mu}\psi$ and $\partial_{\mu}\bar{\psi}$.
To this end, we multiply $\bar{\pi}^{\mu}$ by $\tau_{\mu\nu}$ from the right, and $\pi^{\mu}$ by $\tau_{\nu\mu}$ from the left,
\begin{align}\label{v-dirac}
\pfrac{\bar{\psi}}{x^{\nu}}&=\frac{3m^{*}}{i}\left(\bar{\pi}^{\alpha}-
\frac{i}{2}\bar{\psi}\gamma^{\alpha}\right)\tau_{\alpha\nu}\nonumber\\
\pfrac{\psi}{x^{\nu}}&=\frac{3m^{*}}{i}\,\tau_{\nu\beta}\left(
\pi^{\beta}+\frac{i}{2}\gamma^{\beta}\psi\right).
\end{align}
The Dirac Hamiltonian density is then finally obtained as
\begin{equation}\label{hd-dirac}
\HC_{\text{D}}=
\left(\bar{\pi}^{\alpha}-\frac{i}{2}\bar{\psi}\gamma^{\alpha}
\right)\frac{3m^{*}\tau_{\alpha\beta}}{i}\left(\pi^{\beta}+
\frac{i}{2}\gamma^{\beta}\psi\right)+m\,\bar{\psi}\psi.
\end{equation}
We may expand the products in Eq.~(\ref{hd-dirac}) using Eqs.~(\ref{dirac-algebra}) to find
\begin{equation}\label{hd-dirac2}
\HC_{\text{D}}=im^{*}\left(\frac{1}{2}\bar{\psi}\,\gamma_{\alpha}\pi^{\alpha}-
\frac{1}{2}\bar{\pi}^{\alpha}\gamma_{\alpha}\psi-
3\bar{\pi}^{\alpha}\tau_{\alpha\beta}\pi^{\beta}\right)+
\left(m-m^{*}\right)\bar{\psi}\psi.
\end{equation}
In order to show that the Hamiltonian density $\HC_{\text{D}}$ describes the same dynamics as $\LC_{\text{D}}$ from Eq.~(\ref{ld-dirac}), we set up the canonical equations from Eq.~(\ref{hd-dirac2})
\begin{align*}
\pfrac{\bar{\psi}}{x^{\nu}}&=
\pfrac{\HC_{\text{D}}}{\pi^{\nu}}=\hphantom{-}
im^{*}\left(\onehalf\bar{\psi}\,\gamma_{\nu}-
3\bar{\pi}^{\alpha}\tau_{\alpha\nu}\right)\\
\pfrac{\psi}{x^{\mu}}&=
\pfrac{\HC_{\text{D}}}{\bar{\pi}^{\mu}}=-im^{*}\left(
\onehalf\gamma_{\mu}\psi+3\tau_{\mu\beta}\pi^{\beta}\right).
\end{align*}
Obviously, these equations reproduce the definition of the canonical momenta from Eqs.~(\ref{pi-dirac}) in their inverted form given by Eqs.~(\ref{v-dirac}).
The second set of canonical equations follows from the $\psi$ and $\bar{\psi}$ dependence of the Hamiltonian $\HC_{\text{D}}$,
\begin{align*}
\pfrac{\bar{\pi}^{\alpha}}{x^{\alpha}}=
-\pfrac{\HC_{\text{D}}}{\psi}&=
\frac{im^{*}}{2}\bar{\pi}^{\beta}\gamma_{\beta}-\left(m-m^{*}\right)\bar{\psi}\\
&=\frac{im^{*}}{2}\left(\frac{i}{2}\bar{\psi}\gamma^{\beta}+
\pfrac{\bar{\psi}}{x^{\alpha}}\frac{i\sigma^{\alpha\beta}}{3m^{*}}\right)
\gamma_{\beta}-\left(m-m^{*}\right)\bar{\psi}\\
&=-\frac{i}{2}\pfrac{\bar{\psi}}{x^{\alpha}}\gamma^{\alpha}-m\bar{\psi}\\
\pfrac{\pi^{\alpha}}{x^{\alpha}}=-\pfrac{\HC_{\text{D}}}{\bar{\psi}}&=
-\frac{im^{*}}{2}\gamma_{\beta}\pi^{\beta}-\left(m-m^{*}\right)\psi\\
&=-\frac{im^{*}}{2}\gamma_{\beta}\left(-\frac{i}{2}\gamma^{\beta}\psi+
\frac{i\sigma^{\beta\alpha}}{3m^{*}}\pfrac{\psi}{x^{\alpha}}\right)-\left(m-m^{*}\right)\psi\\
&=\frac{i}{2}\gamma^{\alpha}\pfrac{\psi}{x^{\alpha}}-m\psi.
\end{align*}
The divergences of the canonical momenta follow equally from the derivatives of the first canonical equations, or, equivalently, from the derivatives of Eqs.~(\ref{pi-dirac}),
\begin{align*}
\pfrac{\bar{\pi}^{\alpha}}{x^{\alpha}}&=
\hphantom{-}\frac{i}{2}\pfrac{\bar{\psi}}{x^{\alpha}}\gamma^{\alpha}+
\cancel{\pfrac{^{2}\bar{\psi}}{x^{\alpha}\partial x^{\beta}}\,
\frac{i\sigma^{\alpha\beta}}{3m^{*}}}=\hphantom{-}\frac{i}{2}\pfrac{\bar{\psi}}{x^{\alpha}}\gamma^{\alpha}\\
\pfrac{\pi^{\alpha}}{x^{\alpha}}&=-\frac{i}{2}\gamma^{\alpha}\pfrac{\psi}{x^{\alpha}}-
\cancel{\frac{i\sigma^{\alpha\beta}}{3m^{*}}\,\pfrac{^2\psi}{x^{\alpha}\partial x^{\beta}}}=
-\frac{i}{2}\gamma^{\alpha}\pfrac{\psi}{x^{\alpha}}.
\end{align*}
The terms containing the second derivatives of $\psi$ and $\bar{\psi}$ vanish due to the skew-symmetry of $\sigma^{\mu\nu}$.
Equating finally the expressions for the divergences of the canonical momenta, we encounter, as expected, the Dirac equations~(\ref{el-dirac})
\begin{align*}
\frac{i}{2}\pfrac{\bar{\psi}}{x^{\alpha}}\gamma^{\alpha}&=-m\bar{\psi}-
\frac{i}{2}\pfrac{\bar{\psi}}{x^{\alpha}}\gamma^{\alpha}\\
-\frac{i}{2}\gamma^{\alpha}\pfrac{\psi}{x^{\alpha}}&=
-m\psi+\frac{i}{2}\gamma^{\alpha}\pfrac{\psi}{x^{\alpha}}.
\end{align*}
It should be mentioned that this section is similar to the derivation of the Dirac Hamiltonian density in Ref.~\cite{vonRieth}.
We note that the additional term in the Dirac Lagrangian density $\LC_{\text{D}}^{\prime}$ from Eq.~(\ref{ld-dirac-regular}) --- as compared to the Lagrangian $\LC_{\text{D}}$ from Eq.~(\ref{ld-dirac-symm}) --- entails additional terms in the energy-momentum tensor, namely,
\begin{equation*}
T^{\nu^{\prime}}_{\mu}-T^{\nu}_{\mu}\equiv j_{\mu}^{\nu}(\bx)=\frac{i}{3m^{*}}\left(
\partial_{\alpha}\bar{\psi}\sigma^{\alpha\nu}\partial_{\mu}\psi+
\partial_{\mu}\bar{\psi}\sigma^{\nu\alpha}\partial_{\alpha}\psi-
\delta_{\mu}^{\nu}\partial_{\alpha}\bar{\psi}\sigma^{\alpha\lambda}
\partial_{\lambda}\psi\right).
\end{equation*}
We easily convince ourselves by direct calculation that the divergences of $T^{\nu^{\prime}}_{\mu}$ and $T^{\nu}_{\mu}$ coincide,
\begin{align*}
\pfrac{j_{\mu}^{\beta}}{x^{\beta}}&=\frac{i}{3m^{*}}\Big(
\cancel{\partial_{\beta}\partial_{\alpha}\bar{\psi}\sigma^{\alpha\beta}\partial_{\mu}\psi}+
\partial_{\alpha}\bar{\psi}\sigma^{\alpha\beta}\partial_{\beta}\partial_{\mu}\psi+
\partial_{\beta}\partial_{\mu}\bar{\psi}\sigma^{\beta\alpha}\partial_{\alpha}\psi\\
&\qquad\mbox{}+\cancel{\partial_{\mu}\bar{\psi}\sigma^{\beta\alpha}\partial_{\beta}\partial_{\alpha}\psi}-
\delta_{\mu}^{\beta}\partial_{\beta}\partial_{\alpha}\bar{\psi}\sigma^{\alpha\lambda}\partial_{\lambda}\psi-
\delta_{\mu}^{\beta}\partial_{\alpha}\bar{\psi}\sigma^{\alpha\lambda}\partial_{\beta}\partial_{\lambda}\psi\Big)\\
&=\frac{i}{3m^{*}}\Big(
\partial_{\alpha}\bar{\psi}\sigma^{\alpha\beta}\partial_{\beta}\partial_{\mu}\psi+
\partial_{\beta}\partial_{\mu}\bar{\psi}\sigma^{\beta\alpha}\partial_{\alpha}\psi\\
&\qquad\quad\mbox{}-\partial_{\mu}\partial_{\alpha}\bar{\psi}\sigma^{\alpha\beta}\partial_{\beta}\psi-
\partial_{\alpha}\bar{\psi}\sigma^{\alpha\beta}\partial_{\mu}\partial_{\beta}\psi\Big)\\
&\equiv0,
\end{align*}
which means that the two energy-momentum tensors have the same bulk divergence.
For each index $\mu$, $j_{\mu}^{\nu}(\bx)$ is an identically conserved improvement current associated with the change from $\LC_{\text{D}}$ to $\LC_{\text{D}}^{\prime}$.
The corresponding integrated charges agree when the improvement has no boundary contribution.
\section{\label{examples-ct} Examples of canonical transformations in covariant Hamiltonian field theory}
The formalism of canonical transformations that was worked out in Sect.~\ref{sec:d-w-cantra} is now shown to yield a Hamiltonian representation of Noether's theorem within the admissible class of generators.
Furthermore, a theory of local U$(N)$ gauge transformations is outlined.
\subsection{\label{sec:gen-noether} Hamiltonian Noether theorem for admissible canonical transformations}
Canonical transformations are defined by Eq.~(\ref{varprinzip}) as transformations of the fields $\phi_{I}$ and their conjugate momentum vector fields $\bpi_{I}$ that preserve the variational principle with the corresponding transformed Hamiltonian.
Such a transformation is a {\em symmetry\/} of a specified Hamiltonian system only if the transformed Hamiltonian has the same functional form.
This additional condition leads to a conserved four-current\cite{noether}.
In the following, we work out this correlation by means of an {\em infinitesimal\/} admissible canonical transformation.
The generating function differs from that of the {\em identity\/} transformation by an infinitesimal parameter $\epsilon$ times a characteristic function $g^{\mu}$ according to
\begin{equation}\label{gen-infini}
F_{2}^{\mu}(\phi_{I},\bPi_{I},\bx)=\phi_{J}\,\Pi_{J}^{\mu}+
\epsilon\,g^{\mu}(\phi_{I},\bPi_{I},\bx).
\end{equation}
Admissibility, as established in Sect.~\ref{sec:genf2}, requires
\begin{equation*}
g^{\mu}(\bphi,\bPi,\bx)
=\Pi_{I}^{\mu}\xi_{I}(\bphi,\bx)+b^{\mu}(\bphi,\bx).
\end{equation*}
The functions $\xi_{I}$ and $b^{\mu}$ are independent of the polymomenta.
In particular, $g^{\mu}$ cannot be an arbitrary momentum-dependent conserved current.
At first order in $\epsilon$, replacing $\bPi$ by $\bpi$ in $g^{\mu}$ and its derivatives changes the transformation only at order $\epsilon^{2}$.
With this convention, the subsequent transformation rules follow from the general rules~(\ref{genF2}) as
\begin{align*}
\pi_{I}^{\mu}&=\pfrac{F_{2}^{\mu}}{\phi_{I}}=
\Pi_{I}^{\mu}+\epsilon\,\pfrac{g^{\mu}}{\phi_{I}},\qquad
\Phi_{I}\,\delta^{\mu}_{\nu}=\pfrac{F_{2}^{\mu}}{\Pi_{I}^{\nu}}=
\phi_{I}\,\delta^{\mu}_{\nu}+\epsilon\,\pfrac{g^{\mu}}{\pi_{I}^{\nu}},\\
\HC^{\prime}&=\HC+\left.\pfrac{F_{2}^{\alpha}}{x^{\alpha}}
\right\vert_{\mathrm{expl}}=\HC+\epsilon{\left.\pfrac{g^{\alpha}}
{x^{\alpha}}\right\vert}_{\mathrm{expl}},
\end{align*}
hence
\begin{equation}\label{gl1}
\delta\pi_{I}^{\mu}=-\epsilon\,\pfrac{g^{\mu}}{\phi_{I}},\qquad
\delta\phi_{I}\,\delta^{\mu}_{\nu}=
\epsilon\,\pfrac{g^{\mu}}{\pi_{I}^{\nu}},\qquad
{\delta\HC|}_{\mathrm{CT}}=\epsilon{\left.\pfrac{g^{\alpha}}
{x^{\alpha}}\right\vert}_{\mathrm{expl}}.
\end{equation}
For the admissible form of $g^{\mu}$, the field and momentum variations are, explicitly,
\begin{equation*}
\delta\phi_{I}=\epsilon\xi_{I},\qquad
\delta\pi_{I}^{\mu}=-\epsilon\left(
\pi_{J}^{\mu}\pfrac{\xi_{J}}{\phi_{I}}+
\pfrac{b^{\mu}}{\phi_{I}}\right).
\end{equation*}
As the transformation does not change the independent variables, $x^{\mu}$, both the original as well as the transformed fields refer to the same space-time event $\bx$, hence $\delta x^{\mu}=0$.
Making use of the canonical field equations~(\ref{fgln}), the variation of $\HC$ due to the variations~(\ref{gl1}) of the canonical field variables $\phi_{I}$ and $\pi_{I}^{\mu}$ emerges as
\begin{align}
\delta\HC|_{\mathrm{fields}}&=\pfrac{\HC}{\phi_{I}}\,\delta\phi_{I}+
\pfrac{\HC}{\pi_{I}^{\alpha}}\,\delta\pi_{I}^{\alpha}\nonumber\\
&=-\pfrac{\pi_{I}^{\beta}}{x^{\alpha}}\,\delta_{\beta}^{\alpha}\,
\delta\phi_{I}+\pfrac{\phi_{I}}{x^{\alpha}}\,\delta\pi_{I}^{\alpha}\nonumber\\
&=-\epsilon\left(\pfrac{g^{\alpha}}{\pi_{I}^{\beta}}
\pfrac{\pi_{I}^{\beta}}{x^{\alpha}}+\pfrac{g^{\alpha}}{\phi_{I}}
\pfrac{\phi_{I}}{x^{\alpha}}\right)\nonumber\\
&=-\epsilon\left(\pfrac{g^{\alpha}}{x^{\alpha}}-
{\left.\pfrac{g^{\alpha}}{x^{\alpha}}\right\vert}_{\mathrm{expl}}\right)\nonumber\\
&=-\epsilon\pfrac{g^{\alpha}}{x^{\alpha}}+\delta\HC|_{\mathrm{CT}}.
\label{gl2}
\end{align}
If the infinitesimal transformation rule ${\delta\HC|}_{\mathrm{CT}}$ for the Hamiltonian from Eqs.~(\ref{gl1}) coincides with the variation $\delta\HC|_{\mathrm{fields}}$ at $\delta x^{\mu}=0$ from Eq.~(\ref{gl2}), then the set of infinitesimal transformation rules defines a \emph{symmetry transformation} of the given Hamiltonian system.
The equality of the two Hamiltonian variations must hold as an identity on the phase-space domain.
On using the canonical equations, it is equivalent to
\begin{equation}\label{div-g}
{\delta\HC|}_{\mathrm{CT}}=\delta\HC|_{\mathrm{fields}}
\quad\Longleftrightarrow\quad
\left.\pfrac{g^{\alpha}}{x^{\alpha}}\right|_{\mathrm{can}}=0.
\end{equation}
Here, $|_{\mathrm{can}}$ means substitution of the canonical field equations, not an off-shell zero-divergence condition.
The resulting identity must hold on their full local domain, not merely along one particular solution.
For an admissible characteristic current, the corresponding infinitesimal symmetry transformations therefore satisfy
\begin{align}
\pfrac{g^{\alpha}(\bx)}{x^{\alpha}}&=0\label{div-g1}\\
\delta\pi_{I}^{\mu}=-\epsilon\,\pfrac{g^{\mu}}{\phi_{I}},
\qquad\delta\phi_{I}\,\delta^{\mu}_{\nu}&=
\epsilon\,\pfrac{g^{\mu}}{\pi_{I}^{\nu}},\qquad
\delta\HC=\epsilon{\left.\pfrac{g^{\alpha}}
{x^{\alpha}}\right\vert}_{\mathrm{expl}}.\nonumber
\end{align}
We can now formulate the Hamiltonian Noether theorem and its inverse within the stated class.
\begin{theorem}[Hamiltonian Noether theorem]
Let $F_{2}^{\mu}$ from Eq.~(\ref{gen-infini}) be an admissible infinitesimal generator on the phase-space domain under consideration.
If its transformations preserve the specified Hamiltonian system, i.e., $\delta\HC|_{\mathrm{fields}}=\delta\HC|_{\mathrm{CT}}$ as a phase-space identity, then $g^{\mu}$ is a conserved current on solutions of the canonical equations.

Conversely, if the divergence of an admissible $g^{\mu}$ vanishes as an identity after substitution of the canonical equations on their full local domain, Eq.~(\ref{gl2}) gives the Hamiltonian symmetry condition.
The rules~(\ref{div-g1}) then define the corresponding infinitesimal canonical symmetry.
Boundary conditions are assumed for which the variation of the generating boundary integral vanishes.
\end{theorem}
The converse does not apply to an arbitrary conserved current or to conservation on a single solution.
In the fixed-space-time, derivative-free type-$F_{2}$ class used here, the field transformations are point transformations, as follows from admissibility.
This is a Hamiltonian formulation within that class, not a generalization of Noether's original theorem, which already allowed derivative-dependent transformations\cite{noether}.
\subsubsection{Gauge invariance of the electromagnetic $4$-potential}
For the Maxwell Hamiltonian $\HC_{\mathrm{M}}$ from Eq.~(\ref{hd-maxwell}), the correlation of the $4$-vector potential $a^{\mu}$ with the conjugate fields $p_{\mu\nu}$ is determined by the first field equation~(\ref{fg1-maxwell}) as the generalized curl of $\ba$.
This means on the other hand that the correlation between $\ba$ and the $p_{\mu\nu}$ is {\em not unique}.
Defining a transformed $4$-vector potential $\bA$ according to
\begin{equation}\label{lor-eich}
A_{\mu}=a_{\mu}+\pfrac{\chi(\bx)}{x^{\mu}},
\end{equation}
with $\chi=\chi(\bx)$ an arbitrary smooth function of the independent variables.
This means for the transformation of the $p_{\mu\nu}$
\begin{equation}\label{lor-eich1}
p_{\mu\nu}=\pfrac{a_{\nu}}{x^{\mu}}-\pfrac{a_{\mu}}{x^{\nu}}=
\pfrac{A_{\nu}}{x^{\mu}}-\cancel{\pfrac{^{2}\chi(\bx)}{x^{\nu}\partial x^{\mu}}}-
\pfrac{A_{\mu}}{x^{\nu}}+\cancel{\pfrac{^{2}\chi(\bx)}{x^{\mu}\partial x^{\nu}}}=
P_{\mu\nu}.
\end{equation}
The transformations~(\ref{lor-eich}) and (\ref{lor-eich1}) can be regarded as a canonical transformation.
For the prescribed conserved current $j^{\mu}(\bx)$ of Sect.~\ref{sec:maxwell}, a generating function is
\begin{equation}\label{pt1}
F_{2}^{\mu}(\ba,\bP,\bx)=
\left(a_{\alpha}+\pfrac{\chi(\bx)}{x^{\alpha}}\right)
P^{\alpha\mu}+\chi(\bx)j^{\mu}(\bx).
\end{equation}
This generator depends on the new momenta, but not on their derivatives.
Since the Maxwell momentum is antisymmetric, its derivative is understood as the projected derivative
\begin{equation*}
\left.\pfrac{P^{\alpha\mu}}{P^{\nu\beta}}\right|_{\mathrm{as}}
=\frac{1}{2}\left(\delta_{\nu}^{\alpha}\delta_{\beta}^{\mu}
-\delta_{\beta}^{\alpha}\delta_{\nu}^{\mu}\right).
\end{equation*}
The generating-function rules on this sector are consequently
\begin{equation}\label{genF2b}
\begin{split}
p^{\nu\mu}&=\pfrac{F_{2}^{\mu}}{a_{\nu}},\qquad
\HC^{\prime}=\HC+\left.\pfrac{F_{2}^{\alpha}}
{x^{\alpha}}\right|_{\mathrm{expl}},\\
\frac{1}{2}\left(A_{\nu}\delta_{\beta}^{\mu}
-A_{\beta}\delta_{\nu}^{\mu}\right)
&=\left.\pfrac{F_{2}^{\mu}}{P^{\nu\beta}}\right|_{\mathrm{as}}.
\end{split}
\end{equation}
For Eq.~(\ref{pt1}), these rules yield
\begin{align*}
p^{\nu\mu}&=P^{\nu\mu},\qquad
A_{\nu}=a_{\nu}+\pfrac{\chi}{x^{\nu}},\\
\HC^{\prime}-\HC
&=P^{\alpha\beta}\pfrac{^{2}\chi}{x^{\beta}\partial x^{\alpha}}
+j^{\beta}\pfrac{\chi}{x^{\beta}}
+\chi\pfrac{j^{\beta}}{x^{\beta}}
=j^{\beta}\pfrac{\chi}{x^{\beta}}.
\end{align*}
The first and third terms vanish by antisymmetry and current conservation, respectively.
The remaining term agrees with
\begin{equation*}
\HC_{\mathrm{M}}(\bA,\bP,\bx)
-\HC_{\mathrm{M}}(\ba,\bp,\bx)
=j^{\beta}\pfrac{\chi}{x^{\beta}}.
\end{equation*}
The transformation therefore preserves the functional form of the Maxwell Hamiltonian.
Equations~(\ref{lor-eich}) and (\ref{lor-eich1}) define an ordinary gauge transformation, not a gauge condition.
The {\em Lorenz gauge condition\/} is $\partial_{\mu}a^{\mu}=0$.
Its preservation further requires $\Box\chi=0$.

For the infinitesimal transformation, replace $\chi$ by $\epsilon\chi$.
To first order, the characteristic function is
\begin{equation*}
F_{2}^{\mu}=a_{\alpha}P^{\alpha\mu}
+\epsilon g^{\mu}(\bp,\bx),\qquad
g^{\mu}=p^{\alpha\mu}\pfrac{\chi}{x^{\alpha}}+\chi j^{\mu}.
\end{equation*}
The resulting variations satisfy
\begin{align*}
\delta a_{\nu}&=\epsilon\pfrac{\chi}{x^{\nu}},\qquad
\delta p^{\nu\mu}=0,\\
\delta\HC|_{\mathrm{CT}}
&=\epsilon j^{\alpha}\pfrac{\chi}{x^{\alpha}}
=\pfrac{\HC_{\mathrm{M}}}{a_{\alpha}}\delta a_{\alpha}
=\delta\HC|_{\mathrm{fields}}.
\end{align*}
The Noether current $j_{\mathrm{N}}^{\mu}=g^{\mu}$ has divergence
\begin{align*}
\pfrac{g^{\mu}}{x^{\mu}}
&=\left(\pfrac{p^{\alpha\mu}}{x^{\mu}}+j^{\alpha}\right)
\pfrac{\chi}{x^{\alpha}}
+p^{\alpha\mu}\pfrac{^{2}\chi}{x^{\mu}\partial x^{\alpha}}
+\chi\pfrac{j^{\mu}}{x^{\mu}}\\
&=0
\end{align*}
on solutions of the Maxwell equations.
Using $\partial_{\alpha}p^{\alpha\mu}=j^{\mu}$, we may write this current {\em on shell\/} as
\begin{equation*}
j_{\mathrm{N}}^{\mu}
=\pfrac{}{x^{\alpha}}\left(\chi p^{\alpha\mu}\right).
\end{equation*}
The divergence of this superpotential expression vanishes identically by antisymmetry.
Its integrated charge is a boundary term, and vanishes when $\chi$ has compact spatial support and there are no internal boundaries.
The superpotential expression must not be substituted as an off-shell generator containing derivatives of the new momenta.
\subsection{\label{sec:gen-gauge} General local U$(N)$ gauge transformation}
As an interesting example of a canonical transformation in the covariant Hamiltonian description of classical fields, the general local U$(N)$ gauge transformation is treated in this section.
The main feature of the approach is that the terms to be added to a given Hamiltonian $\HC$ in order to render it {\em locally\/} gauge invariant only depends on the {\em type of fields\/} contained in the Hamiltonian $\HC$ and not on the particular form of the original Hamiltonian itself.
The precondition for this minimal construction is that $\HC$ be invariant under the corresponding {\em global\/} gauge transformation, hence a transformation {\em not\/} depending explicitly on $\bx$.
Additional separately gauge-invariant interactions and alternative gauge-field kinetic terms are not excluded by this requirement.
\subsubsection{External gauge field}
We consider a system consisting of a vector of $N$ complex fields $\phi_{I},\;I=1,\ldots,N$, and the adjoint field vector, $\bar{\bphi}$,
\begin{equation*}
\bphi=\begin{pmatrix}\phi_{1}\\\vdots\\\phi_{N}
\end{pmatrix},\qquad
\bar{\bphi}=\left(\,\bar{\phi}_{1}\cdots\bar{\phi}_{N}\right).
\end{equation*}
A general local linear transformation may be expressed in terms of a dimensionless complex matrix $U(\bx)=(u_{IJ}(\bx))$ and its adjoint, $U^{\dagger}$ that may depend explicitly on the independent variables, $x^{\mu}$, as
\begin{equation}\label{general-pointtra}\begin{split}
\bPhi&=U\,\bphi,\qquad\quad
\bar{\bPhi}=\bar{\bphi}\,U^{\dagger}\\
\Phi_{I}&=u_{IJ}\:\phi_{J},\qquad\,\,\,
\bar{\Phi}_{I}=\bar{\phi}_{J}\,u^{*}_{JI},
\qquad [u_{IJ}]=1.
\end{split}
\end{equation}
Throughout this section, the matrix-entry notation $u^{*}_{IJ}$ means $(U^{\dagger})_{IJ}=\overline{u_{JI}}$, not the entrywise conjugate $\overline{u_{IJ}}$.
An overline denotes entrywise complex conjugation in the matrix identities below.
For scalar fields, $\phi_{I}^{*}=\bar\phi_{I}$ has its usual meaning.
With this notation, $\phi_{I}$ may stand for a set of $I=1,\ldots,N$ complex scalar fields $\phi_{I}$ or Dirac spinors.
In other words, $U$ is supposed to define an isomorphism within the space of the $\phi_{I}$, hence to linearly map the $\phi_{I}$ into objects of the same type.
The uppercase Latin letter indexes label the field or spinor number.
Their transformation in internal space is independent of the space-time metric.
We retain subscripts for these internal labels and distinguish the field representation from its adjoint by bars.
The invariant internal pairing is Hermitian.
In contrast, Greek indexes are used for those components that {\em are\/} associated with a metric --- such as the derivatives with respect to a space-time variable, $x^{\mu}$.
As usual, summation is understood for indexes occurring in pairs.

We restrict ourselves to transformations that preserve the bilinear pairing $\bar{\bphi}\bphi$
\begin{align*}
\bar{\bPhi}\bPhi&=
\bar{\bphi}\,U^{\dagger}U\,\bphi=\bar{\bphi}\bphi
\qquad\qquad\;\;\Longrightarrow\qquad U^{\dagger}U=\Eins=
UU^{\dagger}\\
\bar{\Phi}_{I}\Phi_{I}&=
\bar{\phi}_{J}u^{*}_{JI}\,u_{IK}\phi_{K}=
\bar{\phi}_{K}\phi_{K}\qquad\Longrightarrow\qquad
u^{*}_{JI}\,u_{IK}=\delta_{JK}=u_{JI}\,u^{*}_{IK}.
\end{align*}
This means that $U^{\dagger}=U^{-1}$, hence that the matrix $U$ is supposed to be {\em unitary}.
The transformation~(\ref{general-pointtra}) follows from a generating function that --- corresponding to $\HC$ --- must be a real-valued function of the generally complex fields $\bphi$ and their canonical conjugates, $\bpi^{\mu}$,
\begin{align}
\label{gen-pointtra}
F_{2}^{\mu}(\bphi,\bar{\bphi},\bPi^{\mu},\bar{\bPi}^{\mu},\bx)
&=\bar{\bPi}^{\mu}U\,\bphi+
\bar{\bphi}\,U^{\dagger}\,\bPi^{\mu}\nonumber\\
&=\bar{\Pi}_{K}^{\mu}\,u_{KJ}\,\phi_{J}+
\bar{\phi}_{K}\,u^{*}_{KJ}\,\Pi_{J}^{\mu}.
\end{align}
According to Eqs.~(\ref{genF2}) the set of transformation rules follows as
\begin{align*}
\bar{\pi}_{I}^{\mu}=\pfrac{F_{2}^{\mu}}{\phi_{I}}&=
\bar{\Pi}_{K}^{\mu}u_{KJ}\delta_{IJ},&
\bar{\Phi}_{I}\delta_{\nu}^{\mu}&=
\pfrac{F_{2}^{\mu}}{\Pi_{I}^{\nu}}=\bar{\phi}_{K}
u^{*}_{KJ}\delta_{\nu}^{\mu}\delta_{IJ}\\
\pi_{I}^{\mu}=\pfrac{F_{2}^{\mu}}{\bar{\phi}_{I}}&=
\delta_{IK}u^{*}_{KJ}\Pi_{J}^{\mu},&
\Phi_{I}\delta_{\nu}^{\mu}&=
\pfrac{F_{2}^{\mu}}{\bar{\Pi}_{I}^{\nu}}=
\delta_{\nu}^{\mu}\delta_{IK}u_{KJ}\phi_{J}.
\end{align*}
The complete set of transformation rules and their inverses then read in component notation
\begin{equation}\label{pointtra-rules}\begin{split}
\Phi_{I}&=u_{IJ}\,\phi_{J},\qquad
\bar{\Phi}_{I}=\bar{\phi}_{J}\,u^{*}_{JI},\qquad
\Pi_{I}^{\mu}=u_{IJ}\,\pi_{J}^{\mu},\qquad
\bar{\Pi}_{I}^{\mu}=\bar{\pi}_{J}^{\mu}\,u^{*}_{JI}\\
\phi_{I}&=u^{*}_{IJ}\,\Phi_{J},\qquad
\bar{\phi}_{I}=\bar{\Phi}_{J}u_{JI},\qquad\,
\pi_{I}^{\mu}=u^{*}_{IJ}\,\Pi_{J}^{\mu},\qquad
\bar{\pi}_{I}^{\mu}=\bar{\Pi}_{J}^{\mu}u_{JI}.\end{split}
\end{equation}
We assume the Hamiltonian $\HC$ to be {\em form-invariant\/} under the {\em global\/} gauge transformation~(\ref{general-pointtra}), which is given for $U=\mathrm{const}$, hence for all $u_{IJ}$ {\em not\/} depending on the independent variables, $x^{\mu}$.
In contrast, if $U=U(\bx)$, the transformation~(\ref{pointtra-rules}) is referred to as a {\em local\/} gauge transformation.
The transformation rule for the Hamiltonian is then determined by the explicitly $x^{\mu}$-dependent terms of the generating function $F_{2}^{\mu}$ according to
\begin{align}
\HC^{\prime}-\HC=\left.\pfrac{F_{2}^{\alpha}}{x^{\alpha}}\right\vert
_{\text{expl}}&=
\bar{\Pi}_{I}^{\alpha}\pfrac{u_{IJ}}{x^{\alpha}}\,\phi_{J}+
\bar{\phi}_{I}\pfrac{u^{*}_{IJ}}{x^{\alpha}}\,\Pi_{J}^{\alpha}\nonumber\\
&=\bar{\pi}_{K}^{\alpha}\,u^{*}_{KI}\pfrac{u_{IJ}}{x^{\alpha}}\phi_{J}+
\bar{\phi}_{I}\pfrac{u^{*}_{IJ}}{x^{\alpha}}\,u_{JK}\pi_{K}^{\alpha}\nonumber\\
&=\bar{\pi}_{K}^{\alpha}\,u^{*}_{KI}\pfrac{u_{IJ}}{x^{\alpha}}\phi_{J}+
\bar{\phi}_{K}\pfrac{u^{*}_{KI}}{x^{\alpha}}\,u_{IJ}\pi_{J}^{\alpha}\nonumber\\
&=\left(\bar{\pi}_{K}^{\alpha}\,\phi_{J}-
\bar{\phi}_{K}\pi_{J}^{\alpha}\right)
u^{*}_{KI}\pfrac{u_{IJ}}{x^{\alpha}}.\label{pointtra-ham}
\end{align}
In the last step, the identity
\begin{equation*}
\pfrac{u^{*}_{KI}}{x^{\mu}}\,u_{IJ}+u^{*}_{KI}\,\pfrac{u_{IJ}}{x^{\mu}}=
\pfrac{}{x^{\mu}}\left(u^{*}_{KI}u_{IJ}\right)=\pfrac{}{x^{\mu}}\delta_{KJ}=0
\end{equation*}
was inserted.
If we want to set up a Hamiltonian $\HC_{1}$ that is {\em form-invariant\/} under the {\em local}, hence $x^{\mu}$-dependent transformation generated by~(\ref{gen-pointtra}), then we must compensate the additional terms~(\ref{pointtra-ham}) that emerge from the explicit $x^{\mu}$-dependence of the generating function~(\ref{gen-pointtra}).
A minimal way to achieve this is to {\em adjoin\/} the Hamiltonian $\HC$ of our system with terms that correspond to~(\ref{pointtra-ham}) with regard to their dependence on the canonical variables, $\bphi,\bar{\bphi},\bpi^{\mu},\bar{\bpi}^{\mu}$.
With a {\em unitary\/} matrix $U$, the $u_{IJ}$-dependent terms in Eq.~(\ref{pointtra-ham}) are {\em skew-hermitian},
\begin{equation*}
\overline{u^{*}_{KI}\,\pfrac{u_{IJ}}{x^{\mu}}}=
\pfrac{u^{*}_{JI}}{x^{\mu}}\,u_{IK}=
-u^{*}_{JI}\,\pfrac{u_{IK}}{x^{\mu}},\qquad
\overline{\pfrac{u_{KI}}{x^{\mu}}\,u^{*}_{IJ}}=
u_{JI}\pfrac{u^{*}_{IK}}{x^{\mu}}=
-\pfrac{u_{JI}}{x^{\mu}}u^{*}_{IK},
\end{equation*}
or in matrix notation
\begin{equation*}
{\left(U^{\dagger}\pfrac{U}{x^{\mu}}\right)}^{\dagger}=
\pfrac{U^{\dagger}}{x^{\mu}}U=-U^{\dagger}\pfrac{U}{x^{\mu}},\qquad
{\left(\pfrac{U}{x^{\mu}}U^{\dagger}\right)}^{\dagger}=
U\pfrac{U^{\dagger}}{x^{\mu}}=-\pfrac{U}{x^{\mu}}U^{\dagger}.
\end{equation*}
The $u$-dependent terms in Eq.~(\ref{pointtra-ham}) can thus be compensated by a {\em Hermitian\/} matrix $(\ba_{KJ})$ of ``$4$-vector gauge fields'', with each off-diagonal matrix element, $\ba_{KJ},\;K\neq J$, a complex $4$-vector field with components $a_{KJ\mu},\;\mu=0,\ldots,3$
\begin{equation*}
a_{KJ\mu}=\overline{a_{JK\mu}}.
\end{equation*}
The number of independent gauge fields thus amount to $N^{2}$ real $4$-vectors.
The amended Hamiltonian $\HC_{1}$ thus reads
\begin{equation}\label{tildeHC}
\HC_{1}=\HC+\HC_{\mathrm{a}},\qquad\HC_{\mathrm{a}}=
ig\left(\bar{\pi}_{K}^{\alpha}\phi_{J}-
\bar{\phi}_{K}\pi_{J}^{\alpha}\right)a_{KJ\alpha}.
\end{equation}
With the real coupling constant $g$, the interaction Hamiltonian $\HC_{\mathrm{a}}$ is thus real.
Usually, $g$ is defined to be dimensionless.
We then infer the dimension of the gauge fields $\ba_{KJ}$ to be
\begin{equation*}
[g]=1,\qquad [\ba_{KJ}]=[L]^{-1}=[m]=[\partial_{\mu}].
\end{equation*}
In contrast to the given system Hamiltonian $\HC$, the {\em amended\/} Hamiltonian $\HC_{1}$ is supposed to be {\em invariant in its form\/} under the canonical transformation, hence
\begin{equation}\label{tildeHCp}
\HC_{1}^{\prime}=\HC^{\prime}+\HC_{\mathrm{a}}^{\prime},\qquad
\HC_{\mathrm{a}}^{\prime}=
ig\left(\bar{\Pi}_{K}^{\alpha}\Phi_{J}-
\bar{\Phi}_{K}\Pi_{J}^{\alpha}\right)A_{KJ\alpha}.
\end{equation}
Submitting the amended Hamiltonian $\HC_{1}$ from Eq.~(\ref{tildeHC}) to the canonical transformation generated by Eq.~(\ref{gen-pointtra}), the new Hamiltonian $\HC_{1}^{\prime}$ emerges with Eqs.~(\ref{pointtra-ham}) and (\ref{tildeHCp}) as
\begin{align*}
\HC_{1}^{\prime}&=\HC_{1}+\left.\pfrac{F_{2}^{\alpha}}{x^{\alpha}}
\right\vert_{\text{expl}}=
\HC+\HC_{\mathrm{a}}+\left.\pfrac{F_{2}^{\alpha}}{x^{\alpha}}
\right\vert_{\text{expl}}\\
&=\HC+\left(\bar{\pi}_{K}^{\alpha}\phi_{J}-
\bar{\phi}_{K}\pi_{J}^{\alpha}\right)\left(ig\,a_{KJ\alpha}+
u^{*}_{KI}\pfrac{u_{IJ}}{x^{\alpha}}\right)\\
&\stackrel{!}{=}\HC^{\prime}+\left(\bar{\Pi}_{K}^{\alpha}\Phi_{J}-
\bar{\Phi}_{K}\Pi_{J}^{\alpha}\right)ig\,A_{KJ\alpha}.
\end{align*}
The original base fields, $\phi_{J},\bar{\phi}_{K}$ and their conjugates can now be expressed in terms of the transformed ones according to the rules~(\ref{pointtra-rules}), which yields, after index relabeling, the conditions
\begin{gather*}
\HC^{\prime}(\bPhi,\bar{\bPhi},\bPi^{\mu},\bar{\bPi}^{\mu},x^{\mu})
\stackrel{\text{global GT}}{=}\HC(\bphi,\bar{\bphi},\bpi^{\mu},\bar{\bpi}^{\mu},x^{\mu})\\
\left(\bar{\Pi}_{K}^{\alpha}\Phi_{J}-
\bar{\Phi}_{K}\Pi_{J}^{\alpha}\right)ig\,A_{KJ\alpha}=
\left(\bar{\Pi}_{K}^{\alpha}\Phi_{J}-
\bar{\Phi}_{K}\Pi_{J}^{\alpha}\right)\left(ig\,u_{KL}\,a_{LI\alpha}\,u^{*}_{IJ}+
\pfrac{u_{KI}}{x^{\alpha}}u^{*}_{IJ}\right).
\end{gather*}
This means that the system Hamiltonian must be invariant under the {\em global\/} gauge transformation defined by Eq.~(\ref{pointtra-rules}), whereas the gauge fields $A_{IJ\mu}$ must satisfy the transformation rule
\begin{equation}\label{gauge-tra1}
A_{KJ\mu}=u_{KL}\,a_{LI\mu}\,u^{*}_{IJ}+
\frac{1}{ig}\,\pfrac{u_{KI}}{x^{\mu}}\,u^{*}_{IJ}.
\end{equation}
We observe that for any type of canonical field variables $\phi_{I}$ and for any Hamiltonian system $\HC$, the transformation of the $4$-vector gauge fields $\ba_{IJ}(\bx)$ is uniquely determined according to Eq.~(\ref{gauge-tra1}) by the transformation matrix $U(\bx)$ for the $N$ fields $\phi_{I}$.
In the notation of the $4$-vector gauge fields $\ba_{KJ}(\bx),\;K,J=1,\ldots,N$, the transformation rule is equivalently expressed as
\begin{equation*}
\bA_{KJ}=u_{KL}\,\ba_{LI}\,u^{*}_{IJ}+
\frac{1}{ig}\,\pfrac{u_{KI}}{\bx}\,u^{*}_{IJ},
\end{equation*}
or, in matrix notation
\begin{equation}\label{gauge-tra2}
\hat{A}_{\mu}=U\hat{a}_{\mu}U^{\dagger}+
\frac{1}{ig}\,\pfrac{U}{x^{\mu}}U^{\dagger},\qquad
\hat{\bA}=U\,\hat{\ba}\,U^{\dagger}+
\frac{1}{ig}\,\pfrac{U}{\bx}\,U^{\dagger},
\end{equation}
with $\hat{a}_{\mu}$ denoting the $N\times N$ matrices of the $\mu$-components of the $4$-vectors $\ba_{IK}(\bx)$, and, finally, $\hat{\ba}$ the $N\times N$ matrix of gauge $4$-vectors $\ba_{IK}(\bx)$.
The matrix $U(\bx)$ is {\em unitary}, and thus constitutes a member of the group U$(N)$
\begin{equation*}
U^{\dagger}(\bx)=U^{-1}(\bx),\qquad |\det{U(\bx)}|=1.
\end{equation*}
For $\det{U(\bx)}=+1$, the matrix $U(\bx)$ is a member of the group SU$(N)$.

Inserting the transformation rule for the base fields, $\bPhi=U\,\bphi$, into Eq.~(\ref{gauge-tra2}), we immediately find the {\em homogeneous\/} transformation condition
\begin{equation*}
\pfrac{\bPhi}{x^{\mu}}-ig\,\hat{A}_{\mu}\bPhi=
U\left(\pfrac{\bphi}{x^{\mu}}-ig\,\hat{a}_{\mu}\bphi\right).
\end{equation*}
We identify this ``amended'' partial derivative as the covariant derivative that defines the minimum coupling rule for our gauge transformation.

Equation~(\ref{gauge-tra2}) is the general transformation law for gauge bosons.
$U$ and $\hat{a}_{\mu}$ do not in general commute if $N>1$, hence if $U$ is a unitary matrix rather than a complex number of modulus $1$.
We are then dealing with a non-Abelian gauge theory.
As the matrices $\hat{a}_{\mu}$ are Hermitian, the number of independent gauge $4$-vectors $\ba_{IK}$ amounts to $N$ real vectors on the main diagonal, and $(N^{2}-N)/2$ independent complex off-diagonal vectors, which corresponds to a total number of $N^{2}$ independent real gauge $4$-vectors for a U$(N)$ gauge theory.
Restricting the connection to traceless Hermitian matrices gives $N^{2}-1$ real gauge $4$-vectors for the corresponding SU$(N)$ theory.
\subsubsection{Including the gauge field dynamics}
With the knowledge of the required transformation rule for the gauge fields from Eq.~(\ref{gauge-tra1}), it is now possible to redefine the generating function~(\ref{gen-pointtra}) to also describe the gauge field transformation.
This simultaneously defines the transformation of the canonical conjugates, $p_{JK}^{\mu\nu}$, of the gauge fields $a_{JK\mu}$.
Furthermore, the redefined generating function yields additional terms in the transformation rule for the Hamiltonian.
The additional terms must combine with the original Hamiltonian to satisfy the required canonical transformation law.
They need not be separately invariant.
The transformation rules for the fields $\bphi$ and the gauge field matrices $\hat{\ba}$ (Eq.~(\ref{gauge-tra2})) can be regarded as a canonical transformation that emerges from an explicitly $x^{\mu}$-dependent and real-valued generating function vector of type $F_{2}^{\mu}=F_{2}^{\mu}(\bphi,\bar{\bphi},\bPi,\bar{\bPi},\ba,\bP,\bx)$,
\begin{equation}\label{gen-gaugetra}
F_{2}^{\mu}=\bar{\Pi}_{K}^{\mu}\,u_{KJ}\,\phi_{J}+
\bar{\phi}_{K}\,u^{*}_{KJ}\,\Pi_{J}^{\mu}+P_{JK}^{\alpha\mu}
\left(u_{KL}\,a_{LI\alpha}\,u^{*}_{IJ}+\frac{1}{ig}
\pfrac{u_{KI}}{x^{\alpha}}\,u^{*}_{IJ}\right).
\end{equation}
Accordingly, the subsequent transformation rules for canonical variables $\bphi,\bar{\bphi}$ and their conjugates, $\bpi^{\mu},\bar{\bpi}^{\mu}$, agree with those from Eqs.~(\ref{pointtra-rules}).
The rule for the gauge fields $a_{IK\alpha}$ emerges as
\begin{equation*}
A_{KJ\alpha}\,\delta_{\nu}^{\mu}=
\pfrac{F_{2}^{\mu}}{P_{JK}^{\alpha\nu}}=\delta_{\nu}^{\mu}
\left(u_{KL}\,a_{LI\alpha}\,u^{*}_{IJ}+\frac{1}{ig}
\pfrac{u_{KI}}{x^{\alpha}}u^{*}_{IJ}\right),
\end{equation*}
which obviously coincides with Eq.~(\ref{gauge-tra1}), as demanded.
The transformation of the momentum fields is obtained from the generating function~(\ref{gen-gaugetra}) as
\begin{equation}\label{general-pointtra-gf-deri}
p_{IL}^{\alpha\mu}=\pfrac{F_{2}^{\mu}}{a_{LI\alpha}}=
u^{*}_{IJ}P_{JK}^{\alpha\mu}\,u_{KL}.
\end{equation}
It remains to work out the difference of the Hamiltonians that are submitted to the canonical transformation generated by~(\ref{gen-gaugetra}).
Hence, according to the general rule from Eq.~(\ref{genF2}), we must calculate the divergence of the explicitly $x^{\mu}$-dependent terms of $F_{2}^{\mu}$
\begin{align}
\left.\pfrac{F_{2}^{\alpha}}{x^{\alpha}}\right\vert_{\text{expl}}&=
\bar{\Pi}_{I}^{\alpha}\,\pfrac{u_{IJ}}{x^{\alpha}}\,\phi_{J}+
\bar{\phi}_{I}\,\pfrac{u^{*}_{IJ}}{x^{\alpha}}\,\Pi_{J}^{\alpha}
\label{H-deri-expl}\\
&+P_{JK}^{\alpha\beta}\left(
\pfrac{u_{KL}}{x^{\beta}}a_{LI\alpha}u^{*}_{IJ}+
u_{KL}a_{LI\alpha}\pfrac{u^{*}_{IJ}}{x^{\beta}}+
\frac{1}{ig}\pfrac{u_{KI}}{x^{\alpha}}\pfrac{u^{*}_{IJ}}{x^{\beta}}+
\frac{1}{ig}\pfrac{^{2}u_{KI}}{x^{\alpha}\partial x^{\beta}}u^{*}_{IJ}
\right).\nonumber
\end{align}
We are now going to replace all $u_{IJ}$-dependencies in~(\ref{H-deri-expl}) by canonical variables making use of the canonical transformation rules.
The first two terms on the right-hand side of Eq.~(\ref{H-deri-expl}) can be expressed in terms of the canonical variables by means of the transformation rules~(\ref{pointtra-rules}), (\ref{gauge-tra1}), and (\ref{general-pointtra-gf-deri}) that all follow from the generating function~(\ref{gen-gaugetra})
\begin{align*}
\bar{\Pi}_{I}^{\alpha}\pfrac{u_{IJ}}{x^{\alpha}}\phi_{J}+
\bar{\phi}_{I}\pfrac{u^{*}_{IJ}}{x^{\alpha}}\Pi_{J}^{\alpha}&=
\bar{\Pi}_{I}^{\alpha}\pfrac{u_{IJ}}{x^{\alpha}}u^{*}_{JK}\Phi_{K}+
\bar{\Phi}_{K}u_{KI}\pfrac{u^{*}_{IJ}}{x^{\alpha}}\Pi_{J}^{\alpha}\\
&=\bar{\Pi}_{I}^{\alpha}\pfrac{u_{IJ}}{x^{\alpha}}u^{*}_{JK}\Phi_{K}-
\bar{\Phi}_{K}\pfrac{u_{KI}}{x^{\alpha}}u^{*}_{IJ}\Pi_{J}^{\alpha}\\
&=ig\bar{\Pi}_{I}^{\alpha}\left(A_{IK\alpha}-u_{IL}a_{LJ\alpha}
u^{*}_{JK}\right)\Phi_{K}\\
&\qquad\mbox{}-ig\bar{\Phi}_{K}\left(A_{KJ\alpha}-u_{KL}a_{LI\alpha}
u^{*}_{IJ}\right)\Pi_{J}^{\alpha}\\
&=ig\left(\bar{\Pi}_{K}^{\alpha}\Phi_{J}-
\bar{\Phi}_{K}\Pi_{J}^{\alpha}\right)A_{KJ\alpha}-
ig\left(\bar{\pi}_{K}^{\alpha}\phi_{J}-
\bar{\phi}_{K}\pi_{J}^{\alpha}
\vphantom{\bar{\Pi}_{K}^{\alpha}}\right)a_{KJ\alpha}.
\end{align*}
The second derivative term in Eq.~(\ref{H-deri-expl}) is {\em symmetric\/} in the indexes $\alpha$ and $\beta$.
If we split $P_{JK}^{\alpha\beta}$ into a symmetric $P_{JK}^{(\alpha\beta)}$ and a skew-symmetric part $P_{JK}^{[\alpha\beta]}$ in $\alpha$ and $\beta$
\begin{equation*}
P_{JK}^{\alpha\beta}=P_{JK}^{(\alpha\beta)}+P_{JK}^{[\alpha\beta]},\qquad
P_{JK}^{[\alpha\beta]}=\onehalf\left(
P_{JK}^{\alpha\beta}-P_{JK}^{\beta\alpha}\right),\qquad
P_{JK}^{(\alpha\beta)}=\onehalf\left(
P_{JK}^{\alpha\beta}+P_{JK}^{\beta\alpha}\right),
\end{equation*}
then the second derivative term vanishes for $P_{JK}^{[\alpha\beta]}$,
\begin{equation*}
P_{JK}^{[\alpha\beta]}\pfrac{^{2}u_{KI}}{x^{\alpha}\partial x^{\beta}}=0.
\end{equation*}
By inserting the transformation rules for the gauge fields from Eqs.~(\ref{gauge-tra1}), the remaining terms of (\ref{H-deri-expl}) for the skew-symmetric part of $P_{JK}^{\alpha\beta}$ are converted into
\begin{align*}
&\quad\,\,P_{JK}^{[\alpha\beta]}\left(
\pfrac{u_{KL}}{x^{\beta}}\,a_{LI\alpha}\,u^{*}_{IJ}+
u_{KL}\,a_{LI\alpha}\,\pfrac{u^{*}_{IJ}}{x^{\beta}}+
\frac{1}{ig}\pfrac{u_{KI}}{x^{\alpha}}\pfrac{u^{*}_{IJ}}{x^{\beta}}\right)\\
&=ig\,p_{JK}^{[\alpha\beta]}\,a_{KI\alpha}\,a_{IJ\beta}-
ig\,P_{JK}^{[\alpha\beta]}\,A_{KI\alpha}\,A_{IJ\beta}\\
&=\onehalf ig\left(p_{JK}^{\alpha\beta}-
p_{JK}^{\beta\alpha}\right)a_{KI\alpha}\,a_{IJ\beta}-
\onehalf ig\left(P_{JK}^{\alpha\beta}-
P_{JK}^{\beta\alpha}\right)A_{KI\alpha}\,A_{IJ\beta}\\
&=\onehalf ig\,p_{JK}^{\alpha\beta}\left(a_{KI\alpha}\,
a_{IJ\beta}-a_{KI\beta}\,a_{IJ\alpha}\right)-
\onehalf ig\,P_{JK}^{\alpha\beta}\left(A_{KI\alpha}\,
A_{IJ\beta}-A_{KI\beta}\,A_{IJ\alpha}\right).
\end{align*}
For the symmetric part of $P_{JK}^{\alpha\beta}$, we obtain
\begin{align*}
&\quad\,\,P_{JK}^{(\alpha\beta)}\left(
\pfrac{u_{KL}}{x^{\beta}}\,a_{LI\alpha}
\,u^{*}_{IJ}+u_{KL}a_{LI\alpha}
\,\pfrac{u^{*}_{IJ}}{x^{\beta}}+
\frac{1}{ig}\pfrac{u_{KI}}{x^{\alpha}}
\,\pfrac{u^{*}_{IJ}}{x^{\beta}}+\frac{1}{ig}
\pfrac{^{2}u_{KI}}{x^{\alpha}\partial x^{\beta}}\,
u^{*}_{IJ}\right)\\
&=P_{JK}^{(\alpha\beta)}\left(
\pfrac{A_{KJ\alpha}}{x^{\beta}}-u_{KL}\,
\pfrac{a_{LI\alpha}}{x^{\beta}}\,u^{*}_{IJ}\right)\\
&=\onehalf P_{JK}^{\alpha\beta}\left(
\pfrac{A_{KJ\alpha}}{x^{\beta}}+\pfrac{A_{KJ\beta}}{x^{\alpha}}\right)-
\onehalf p_{JK}^{\alpha\beta}\left(\pfrac{a_{KJ\alpha}}{x^{\beta}}+
\pfrac{a_{KJ\beta}}{x^{\alpha}}\right).
\end{align*}
In summary, by inserting the transformation rules into Eq.~(\ref{H-deri-expl}), the divergence of the explicitly $x^{\mu}$-dependent terms of $F_{2}^{\mu}$ --- and hence the difference of transformed and original Hamiltonians --- can be expressed completely in terms of the canonical variables as
\begin{align*}
\left.\pfrac{F_{2}^{\alpha}}{x^{\alpha}}\right\vert_{\text{expl}}&=
ig\,\Big[\left(\bar{\Pi}_{K}^{\alpha}\Phi_{J}-
\bar{\Phi}_{K}\Pi_{J}^{\alpha}\right)A_{KJ\alpha}-
\left(\vphantom{\bar{\Pi}_{K}^{\alpha}}
\bar{\pi}_{K}^{\alpha}\phi_{J}-
\bar{\phi}_{K}\pi_{J}^{\alpha}\right)a_{KJ\alpha}\\
&\quad\mbox{}-\onehalf P_{JK}^{\alpha\beta}\left(
A_{KI\alpha}\,A_{IJ\beta}-A_{KI\beta}\,A_{IJ\alpha}\right)+
\onehalf p_{JK}^{\alpha\beta}\left(
a_{KI\alpha}\,a_{IJ\beta}-a_{KI\beta}\,a_{IJ\alpha}\right)\Big]\\
&\quad\mbox{}+\onehalf P_{JK}^{\alpha\beta}\left(
\pfrac{A_{KJ\alpha}}{x^{\beta}}+\pfrac{A_{KJ\beta}}{x^{\alpha}}\right)-
\onehalf p_{JK}^{\alpha\beta}\left(\pfrac{a_{KJ\alpha}}{x^{\beta}}+
\pfrac{a_{KJ\beta}}{x^{\alpha}}\right).
\end{align*}
We observe that {\em all\/} $u_{IJ}$-dependencies of Eq.~(\ref{H-deri-expl}) were expressed {\em symmetrically\/} in terms of the original and transformed complex scalar fields $\phi_{J},\Phi_{J}$ and $4$-vector gauge fields $\ba_{JK},\bA_{JK}$, in conjunction with their respective canonical momenta.
The intermediate expressions $\HC_{2}$ and $\HC_{3}$ below contain derivatives of the gauge fields.
They are therefore not ordinary phase-space Hamiltonians of the form introduced in Sect.~\ref{sec:d-w-theory}.
Their role is to specify the first-order action.
The symmetric derivative terms cancel there, as shown in Eqs.~(\ref{varprinzip2})--(\ref{H-tilde2}), leaving the derivative-free expression $\HC_{4}$.
With this understanding, an amended Hamiltonian expression $\HC_{2}$ of the form
\begin{align}
\HC_{2}&=
\HC(\bpi,\bphi,\bx)+ig\left(\bar{\pi}_{K}^{\alpha}\phi_{J}-
\bar{\phi}_{K}\pi_{J}^{\alpha}\right)a_{KJ\alpha}\nonumber\\
&\quad\mbox{}-\onehalf p_{JK}^{\alpha\beta}\left[ig \left(
a_{KI\alpha}\,a_{IJ\beta}-a_{KI\beta}\,a_{IJ\alpha}\right)-
\pfrac{a_{KJ\alpha}}{x^{\beta}}-\pfrac{a_{KJ\beta}}{x^{\alpha}}\right]
\label{amended-H}
\end{align}
is then transformed according to the general rule~(\ref{genF2})
\begin{equation*}
\HC_{2}^{\prime}=\HC_{2}+{\left.\pfrac{F_{2}^{\alpha}}
{x^{\alpha}}\right\vert}_{\text{expl}}
\end{equation*}
into the new Hamiltonian
\begin{align}
\HC_{2}^{\prime}&=
\HC(\bPi,\bPhi,\bx)+ig\left(\bar{\Pi}_{K}^{\alpha}\Phi_{J}-
\bar{\Phi}_{K}\Pi_{J}^{\alpha}\right)A_{KJ\alpha}\nonumber\\
&\quad\mbox{}-\onehalf P_{JK}^{\alpha\beta}\left[ig\left(
A_{KI\alpha}\,A_{IJ\beta}-A_{KI\beta}\,A_{IJ\alpha}\right)-
\pfrac{A_{KJ\alpha}}{x^{\beta}}-\pfrac{A_{KJ\beta}}{x^{\alpha}}\right].
\label{amended-Hp}
\end{align}
The entire transformation is thus {\em form-conserving\/} provided that the original Hamiltonian $\HC(\bpi,\bphi,\bx)$ is also form-invariant if expressed in terms of the new fields, $\HC(\bPi,\bPhi,\bx)$, according to the transformation rules~(\ref{pointtra-rules}).
In other words, $\HC(\bpi,\bphi,\bx)$ must be form-invariant under the corresponding {\em global\/} gauge transformation.

In order for the presented transformation theory to be {\em physically consistent}, we must ensure that the {\em canonical field equations\/} for the derivatives of the gauge fields that follow from the final form-invariant amended Hamiltonians, $\HC_{3}$ and $\HC_{3}^{\prime}$, coincide with the derivatives of the transformation rules for the gauge fields from Eq.~(\ref{gauge-tra1}).
To supply nontrivial gauge-field kinetic dynamics, we further amend the form-invariant expressions $\HC_{2}$ from Eq.~(\ref{amended-H}) and $\HC_{2}^{\prime}$ from Eq.~(\ref{amended-Hp}) by invariant terms $\HC_{\text{dyn}}(\bp)$ and $\HC_{\text{dyn}}^{\prime}(\bP)$, respectively, according to
\begin{align*}
\HC_{3}^{\prime}&=\HC(\bPi,\bPhi,\bx)+\HC_{\text{dyn}}^{\prime}(\bP)+
ig\left(\bar{\Pi}_{K}^{\alpha}\Phi_{J}-
\bar{\Phi}_{K}\Pi_{J}^{\alpha}\right)A_{KJ\alpha}\nonumber\\
&\quad\mbox{}-\onehalf ig P_{JK}^{\alpha\beta}\left(
A_{KI\alpha}\,A_{IJ\beta}-A_{KI\beta}\,A_{IJ\alpha}\right)+
\onehalf P_{JK}^{\alpha\beta}\left(\pfrac{A_{KJ\alpha}}{x^{\beta}}+
\pfrac{A_{KJ\beta}}{x^{\alpha}}\right).
\end{align*}
Of course, $\HC_{\text{dyn}}^{\prime}(\bP)$ must be form-invariant as well in order to ensure the form-invariance of the {\em final\/} amended Hamiltonians, $\HC_{3}$ and $\HC_{3}^{\prime}$.
Varying the first-order action with respect to $P_{JK}^{\mu\nu}$, with the field derivatives held fixed during that variation, gives the relation
\begin{equation*}
\pfrac{A_{KJ\mu}}{x^{\nu}}=\pfrac{\HC_{3}^{\prime}}{P_{JK}^{\mu\nu}}=
\pfrac{\HC_{\text{dyn}}^{\prime}}{P_{JK}^{\mu\nu}}-\onehalf ig\left(
A_{KI\mu}\,A_{IJ\nu}-A_{KI\nu}\,A_{IJ\mu}\right)+\onehalf\left(
\pfrac{A_{KJ\mu}}{x^{\nu}}+\pfrac{A_{KJ\nu}}{x^{\mu}}\right).
\end{equation*}
Applying now the transformation rules~(\ref{gauge-tra1}), for the gauge fields $\bA_{KJ}$, we find after straightforward calculation
\begin{align*}
\pfrac{\HC_{\text{dyn}}^{\prime}}{P_{JK}^{\mu\nu}}&=
\onehalf\left(\pfrac{A_{KJ\mu}}{x^{\nu}}-\pfrac{A_{KJ\nu}}{x^{\mu}}\right)+
\onehalf ig\left(A_{KI\mu}\,A_{IJ\nu}-A_{KI\nu}\,A_{IJ\mu}\right)\\
&=\onehalf u_{KL}\left[\pfrac{a_{LN\mu}}{x^{\nu}}-\pfrac{a_{LN\nu}}{x^{\mu}}+
ig\left(a_{LI\mu}\,a_{IN\nu}-a_{LI\nu}\,a_{IN\mu}\right)\right]u^{*}_{NJ}\\
&=u_{KL}\,\pfrac{\HC_{\text{dyn}}}{p_{NL}^{\mu\nu}}\,u^{*}_{NJ}.
\end{align*}
The derivatives of $\HC_{\text{dyn}}$ and $\HC_{\text{dyn}}^{\prime}$ obviously transform like the canonical momenta, as stated in Eq.~(\ref{general-pointtra-gf-deri}).
Covariance alone does not identify quantities with the same transformation law.
We choose the conventional quadratic kinetic term, for which the momentum-field-strength relation is linear with the normalization
\begin{equation*}
\pfrac{\HC_{\text{dyn}}^{\prime}}{P_{JK}^{\mu\nu}}=-\onehalf P_{KJ\mu\nu},\qquad
\pfrac{\HC_{\text{dyn}}}{p_{JK}^{\mu\nu}}=-\onehalf p_{KJ\mu\nu}.
\end{equation*}
This means, in turn, that $\HC_{\text{dyn}}^{\prime}$ and thus $\HC_{\text{dyn}}$ are given by
\begin{equation}\label{ptimesp-trans}
\HC_{\text{dyn}}^{\prime}(\bP)=-\quarter P_{JK}^{\alpha\beta}P_{KJ\alpha\beta},\qquad
\HC_{\text{dyn}}(\bp)=-\quarter p_{JK}^{\alpha\beta}p_{KJ\alpha\beta}.
\end{equation}
Equation~(\ref{ptimesp-trans}) is the simplest quadratic, Lorentz- and U$(N)$-invariant choice with the conventional normalization, not the only consistent choice.
For example,
\begin{equation*}
X=p_{JK}^{\alpha\beta}p_{KJ\alpha\beta},\qquad
\HC_{\text{dyn}}=-\frac14 X+\lambda X^{2}
\end{equation*}
is also invariant, for a constant $\lambda$ of the appropriate dimension, and its derivative
\begin{equation*}
\pfrac{\HC_{\text{dyn}}}{p_{JK}^{\mu\nu}}
=\left(-\frac12+4\lambda X\right)p_{KJ\mu\nu}
\end{equation*}
has the required covariance.
It gives a different constitutive relation between momentum and field strength.
Even among quadratic U$(N)$ invariants, a term proportional to $(\operatorname{tr}p^{\alpha\beta}) (\operatorname{tr}p_{\alpha\beta})$ allows an independent weight for the trace sector.
In the following we retain Eq.~(\ref{ptimesp-trans}).
With this choice, the amended Hamiltonian expression $\HC_{3}$ is given by
\begin{align}
\HC_{3}=&\HC+\HC_{\mathrm{g}}\label{H-tilde}\\
\HC_{\mathrm{g}}=&ig\left(\bar{\pi}_{K}^{\alpha}\phi_{J}-
\bar{\phi}_{K}\pi_{J}^{\alpha}\right)\,a_{KJ\alpha}-
\quarter p_{JK}^{\alpha\beta}\,p_{KJ\alpha\beta}\nonumber\\
&-\onehalf p_{JK}^{\alpha\beta}\left[ig\,\left(
a_{KI\alpha}\,a_{IJ\beta}-a_{KI\beta}\,a_{IJ\alpha}\right)-
\pfrac{a_{KJ\alpha}}{x^{\beta}}-\pfrac{a_{KJ\beta}}{x^{\alpha}}\right].\nonumber
\end{align}
%
\subsubsection{Inserting the gauge-invariant Hamiltonian $\HC_{3}$ into the action integral}
With gauge fields $a_{KJ\mu}$ and their conjugates, $p_{JK}^{\mu\nu}$, the additional dynamical quantities of the locally gauge-invariant system, the amended action integral from Eq.~(\ref{varprinzip}) reads
\begin{equation}\label{varprinzip2}
S=\int_{R}\left(\bar\pi_{I}^{\beta}\pfrac{\phi_{I}}{x^{\beta}}+
\pfrac{\bar\phi_{I}}{x^{\beta}}\pi_{I}^{\beta}+
p_{JK}^{\alpha\beta}\pfrac{a_{KJ\alpha}}{x^{\beta}}-\HC_{3}\right)d^{4}x.
\end{equation}
Inserting the explicit representation of $\HC_{3}$ from Eq.~(\ref{H-tilde}) then yields the following non-standard form of the action integral
\begin{equation}\label{varprinzip3}
S=\int_{R}\Bigg[\bar\pi_{I}^{\beta}\pfrac{\phi_{I}}{x^{\beta}}+
\pfrac{\bar\phi_{I}}{x^{\beta}}\pi_{I}^{\beta}
+\onehalf p_{JK}^{\alpha\beta}\left(\pfrac{a_{KJ\alpha}}{x^{\beta}}-
\pfrac{a_{KJ\beta}}{x^{\alpha}}\right)-\HC_{4}\Bigg]d^{4}x,
\end{equation}
with
\begin{align}
\HC_{4}=\HC&+ig\left(\bar{\pi}_{K}^{\alpha}\phi_{J}-
\bar{\phi}_{K}\pi_{J}^{\alpha}\right)\,a_{KJ\alpha}-
\quarter p_{JK}^{\alpha\beta}\,p_{KJ\alpha\beta}\nonumber\\
&-\onehalf ig\,p_{JK}^{\alpha\beta}\left(
a_{KI\alpha}\,a_{IJ\beta}-a_{KI\beta}\,a_{IJ\alpha}\right)
\label{H-tilde2}
\end{align}
In Eq.~(\ref{varprinzip3}), the derivative and commutator couplings involve only $p_{JK}^{[\alpha\beta]}$.
The quadratic term in $\HC_{4}$ also contains the symmetric part, as shown by
\begin{equation*}
p_{JK}^{\alpha\beta}p_{KJ\alpha\beta}
=p_{JK}^{[\alpha\beta]}p_{KJ[\alpha\beta]}
+p_{JK}^{(\alpha\beta)}p_{KJ(\alpha\beta)}.
\end{equation*}
If all components of $p_{JK}^{\alpha\beta}$ are varied independently, variation of the symmetric part of the action yields $p_{JK}^{(\alpha\beta)}=0$.
Thus it is an algebraic auxiliary variable that vanishes on shell, not a part absent from the action.
After eliminating it, the remaining variations are antisymmetric.
Equivalently, one may formulate the reduced theory in the antisymmetric sector from the outset.
The full action combines the matter terms into covariant derivatives and the antisymmetric gauge terms into the field strength.
The quadratic auxiliary term is invariant separately.
It is therefore {\em form-invariant\/} under the local U$(N)$ transformations.
The canonical equation for the derivative of the gauge fields is now obtained directly from~(\ref{varprinzip3}) as
\begin{equation}\label{caneq-p}
\frac{1}{2}\left(\pfrac{a_{KJ\alpha}}{x^{\beta}}-
\pfrac{a_{KJ\beta}}{x^{\alpha}}\right)=\pfrac{\HC_{4}}{p_{JK}^{\alpha\beta}}.
\end{equation}
With $\HC_{4}$ from Eq.~(\ref{H-tilde2}), this reads in explicit form
\begin{align*}
\frac{1}{2}\left(\pfrac{a_{KJ\alpha}}{x^{\beta}}-
\pfrac{a_{KJ\beta}}{x^{\alpha}}\right)=-\onehalf p_{KJ\alpha\beta}-
\onehalf ig\left(a_{KI\alpha}\,a_{IJ\beta}-a_{KI\beta}\,a_{IJ\alpha}\right),
\end{align*}
hence
\begin{equation}\label{can-momentum-gf}
p_{KJ\mu\nu}=\pfrac{a_{KJ\nu}}{x^{\mu}}-\pfrac{a_{KJ\mu}}{x^{\nu}}+
ig\left(a_{KI\nu}\,a_{IJ\mu}-a_{KI\mu}\,a_{IJ\nu}\right).
\end{equation}
We observe that $p_{KJ\mu\nu}$ is skew-symmetric in the indices $\mu,\nu$.
For the chosen action with unrestricted momentum variation, this follows on shell and includes the algebraic equation for the symmetric part.
After its elimination, the reduced action and antisymmetric variation prescription must accompany the final Hamiltonian $\HC_{4}$ given by
\begin{align}
\HC_{4}&=\HC+\HC_{\mathrm{g}},\qquad
p_{JK}^{\mu\nu}=-p_{JK}^{\nu\mu}\nonumber\\
\HC_{\mathrm{g}}&=-\quarter p_{JK}^{\alpha\beta}\,p_{KJ\alpha\beta}+
ig\left(\bar{\pi}_{K}^{\alpha}\,a_{KJ\alpha}\,\phi_{J}-
\bar{\phi}_{K}\,a_{KJ\alpha}\,\pi_{J}^{\alpha}-
p_{JK}^{\alpha\beta}\,a_{KI\alpha}\,a_{IJ\beta}\right).
\label{H-g2}
\end{align}
Thus, $\HC_{\mathrm{g}}$ describes the dynamics of {\em massless\/} $4$-vector fields $\ba_{IK}$, namely, their couplings to the base fields $\phi_{I}$ as well as their self-couplings.
This is the result of the local U$(N)$ construction with the chosen quadratic gauge kinetic term and no additional separately gauge-invariant interactions.

From the locally gauge-invariant Hamiltonian~(\ref{H-g2}), the canonical equation for the base fields $\phi_{I}$ is given by
\begin{align*}
{\left.\pfrac{\phi_{I}}{x^{\mu}}\right|}_{\HC_{4}}&=
\pfrac{\HC_{4}}{\bar{\pi}_{I}^{\mu}}=
\pfrac{\HC}{\bar{\pi}_{I}^{\mu}}+ig\,a_{IJ\mu}\phi_{J}\\
&={\left.\pfrac{\phi_{I}}{x^{\mu}}\right|}_{\HC}+ig\,a_{IJ\mu}\phi_{J}.
\end{align*}
This is exactly the so-called ``minimum coupling rule'', which is also referred to as the ``gauge covariant derivative''.
Remarkably, in the canonical formalism this result is {\em derived}, hence does not need to be postulated.
It is commonly assumed that the quantities $a_{IJ\mu}$ exhibit \emph{elementary} fields themselves, hence that the $a_{IJ\mu}$ are not compositions of elementary fields.
\subsection{Locally gauge-invariant Lagrangian}
\subsubsection{Legendre transformation for a general system Hamiltonian}
The equivalent gauge-invariant Lagrangian $\LC_{3}$ is obtained by eliminating the canonical momenta from the first-order action, using $\HC_{3}$ from Eq.~(\ref{H-tilde}).
In the matter sector this is the inverse Legendre transformation.
In the gauge sector it is the restricted elimination described above.
This yields
\begin{equation*}
\LC_{3}=\bar{\pi}_{K}^{\alpha}\pfrac{\phi_{K}}{x^{\alpha}}+
\pfrac{\bar{\phi}_{K}}{x^{\alpha}}\pi_{K}^{\alpha}+
p_{JK}^{\alpha\beta}\pfrac{a_{KJ\alpha}}{x^{\beta}}-\HC_{3},\qquad
\HC_{3}=\HC+\HC_{\mathrm{g}}.
\end{equation*}
With $p_{JK}^{\mu\nu}$ from Eq.~(\ref{can-momentum-gf}) and $\HC_{\mathrm{g}}$ from Eq.~({\ref{H-tilde}}), we thus have
\begin{align*}
p_{JK}^{\alpha\beta}\pfrac{a_{KJ\alpha}}{x^{\beta}}-\HC_{\mathrm{g}}&=
\onehalf p_{JK}^{\alpha\beta}\left(\pfrac{a_{KJ\alpha}}{x^{\beta}}-
\pfrac{a_{KJ\beta}}{x^{\alpha}}\right)+
\onehalf p_{JK}^{\alpha\beta}\left(\pfrac{a_{KJ\alpha}}{x^{\beta}}+
\pfrac{a_{KJ\beta}}{x^{\alpha}}\right)-\HC_{\mathrm{g}}\\
&=-\onehalf p_{JK}^{\alpha\beta}\,p_{KJ\alpha\beta}-
\onehalf ig\,p_{JK}^{\alpha\beta}\left(a_{KI\alpha}\,a_{IJ\beta}-
a_{KI\beta}\,a_{IJ\alpha}\right)\\
&\quad\mbox{}+
\onehalf p_{JK}^{\alpha\beta}\left(\pfrac{a_{KJ\alpha}}{x^{\beta}}+
\pfrac{a_{KJ\beta}}{x^{\alpha}}\right)-\HC_{\mathrm{g}}\\
&=-ig\left(\bar{\pi}_{K}^{\alpha}\phi_{J}-
\bar{\phi}_{K}\pi_{J}^{\alpha}\right)\,a_{KJ\alpha}-
\quarter p_{JK}^{\alpha\beta}\,p_{KJ\alpha\beta}.
\end{align*}
The locally gauge-invariant Lagrangian $\LC_{3}$ for any given globally gauge-invariant system Hamiltonian $\HC(\bar{\phi}_{I},\allowbreak\phi_{I},\allowbreak \bar{\bpi}_{I},\allowbreak\bpi_{I},\allowbreak x)$ is then
\begin{align}\label{general-invariant-lagrangian}
\LC_{3}&=-\quarter p_{JK}^{\alpha\beta}\,p_{KJ\alpha\beta}-
ig\left(\bar{\pi}_{K}^{\alpha}\phi_{J}-
\bar{\phi}_{K}\pi_{J}^{\alpha}\right)a_{KJ\alpha}+
\bar{\pi}_{K}^{\alpha}\pfrac{\phi_{K}}{x^{\alpha}}+
\pfrac{\bar{\phi}_{K}}{x^{\alpha}}\pi_{K}^{\alpha}-\HC\\
&=-\quarter p_{JK}^{\alpha\beta}\,p_{KJ\alpha\beta}+
\bar{\pi}_{K}^{\alpha}\left(\pfrac{\phi_{K}}{x^{\alpha}}-
ig\,a_{KJ\alpha}\phi_{J}\right)+\left(\pfrac{\bar{\phi}_{K}}{x^{\alpha}}+
ig\,\bar{\phi}_{J}a_{JK\alpha}\right)\pi_{K}^{\alpha}-\HC.\nonumber
\end{align}
As implied by the Lagrangian formalism, the dynamical variables are given by both the fields, $\bar{\phi}_{K}$, $\phi_{J}$, and $a_{KJ\alpha}$, in conjunction with their respective partial derivatives with respect to the independent variables, $x^{\mu}$.
Therefore, the $\bp_{KJ}$ in $\LC_{3}$ from Eq.~(\ref{general-invariant-lagrangian}) are now merely abbreviations for a combination of the Lagrangian dynamical variables.
For the quadratic gauge kinetic term chosen here, independently of the given matter Hamiltonian $\HC$, the relation of $\bp_{KJ}$ to the gauge fields and their derivatives is Eq.~(\ref{can-momentum-gf}).
Other invariant kinetic terms need not give this same constitutive relation.

The correlation of the momenta $\bpi_{I},\bar{\bpi}_{I}$ to the base fields $\phi_{I},\bar{\phi}_{I}$ and their derivatives are derived from Eq.~(\ref{general-invariant-lagrangian}) for the given system Hamiltonian $\HC$ via
\begin{equation}\label{pi-phip}
\pfrac{\HC}{\bar{\pi}_{I}^{\mu}}=\pfrac{\phi_{I}}{x^{\mu}}-
ig\,a_{IJ\mu}\phi_{J},\qquad
\pfrac{\HC}{\pi_{I}^{\mu}}=\pfrac{\bar{\phi}_{I}}{x^{\mu}}+
ig\,\bar{\phi}_{J}\,a_{JI\mu}.
\end{equation}
Thus, for any {\em globally\/} gauge-invariant system Hamiltonian $\HC(\bar{\phi}_{I},\phi_{I},\bar{\bpi}_{I},\bpi_{I},x)$, the amen\-ded Lagrangian $\LC_{3}$ from Eq.~(\ref{general-invariant-lagrangian}) with the $\bar{\bpi}_{I},\bpi_{I}$ to be determined from Eqs.~(\ref{pi-phip}) describes in the Lagrangian formalism the associated physical system that is invariant under {\em local\/} gauge transformations.
\subsubsection{Klein-Gordon system Hamiltonian}
The generalized Klein-Gordon Hamiltonian $\HC_{\text{KG}}$ describing $N$ complex scalar fields $\phi_{I}$ that are associated with equal masses $m$ is
\begin{equation*}
\HC_{\text{KG}}(\bpi_{\mu},\bpi^{*\,\mu},\bphi,\bphi^{*})=
\pi_{I\alpha}^{*}\pi_{I}^{\alpha}+m^{2}\,\phi_{I}^{*}\phi_{I}.
\end{equation*}
This Hamiltonian is clearly form-invariant under the global gauge-transformation defined by Eqs.~(\ref{pointtra-rules}).
Following Eqs.~(\ref{H-tilde}) and (\ref{H-g2}), the corresponding locally gauge-invariant Hamiltonian $\HC_{3,\text{KG}}$ is then
\begin{align*}
\HC_{3,\mathrm{KG}}&=\pi_{I\alpha}^{*}\pi_{I}^{\alpha}+
m^{2}\,\phi_{I}^{*}\phi_{I}-\quarter p_{JK}^{\alpha\beta}\,p_{KJ\alpha\beta}\\
&\quad\mbox{}+ig\left(\pi_{K}^{*\,\alpha}\,a_{KJ\alpha}\,\phi_{J}-
\phi_{K}^{*}\,a_{KJ\alpha}\,\pi_{J}^{\alpha}-
p_{JK}^{\alpha\beta}\,a_{KI\alpha}\,a_{IJ\beta}\right),\qquad
p_{JK}^{\mu\nu}&\stackrel{!}{=}-p_{JK}^{\nu\mu}.
\end{align*}
To derive the equivalent locally gauge-invariant Lagrangian $\LC_{3,\mathrm{KG}}$, we set up the first canonical equation for the gauge-invariant Hamiltonian $\HC_{3,\text{KG}}$ of our actual example
\begin{equation*}
\pfrac{\phi_{I}}{x^{\mu}}=\pfrac{\HC_{3,\text{KG}}}{\pi_{I}^{*\,\mu}}=
\pi_{I\mu}+ig\,a_{IJ\mu}\phi_{J},\qquad
\pfrac{\phi_{I}^{*}}{x^{\mu}}=\pfrac{\HC_{3,\text{KG}}}{\pi_{I}^{\mu}}=
\pi_{I\mu}^{*}-ig\,\phi_{J}^{*}\,a_{JI\mu}.
\end{equation*}
Inserting $\partial\phi_{I}/\partial x^{\mu}$ and $\partial\phi_{I}^{*}/\partial x^{\mu}$ into Eq.~(\ref{general-invariant-lagrangian}), we directly encounter the {\em locally\/} gauge-invariant Lagrangian $\LC_{3,\text{KG}}$ as
\begin{equation*}
\LC_{3,\text{KG}}=\pi_{I\alpha}^{*}\pi_{I}^{\alpha}-
m^{2}\,\phi_{I}^{*}\phi_{I}-\quarter p_{JK}^{\alpha\beta}\,p_{KJ\alpha\beta},
\end{equation*}
with the abbreviations
\begin{align*}
\pi_{I\mu}&=\pfrac{\phi_{I}}{x^{\mu}}-ig\,a_{IJ\mu}\phi_{J},\qquad
\pi_{I\mu}^{*}=\pfrac{\phi_{I}^{*}}{x^{\mu}}+ig\,\phi_{J}^{*}\,a_{JI\mu}\\
p_{KJ\mu\nu}&=\pfrac{a_{KJ\nu}}{x^{\mu}}-\pfrac{a_{KJ\mu}}{x^{\nu}}+
ig\left(a_{KI\nu}\,a_{IJ\mu}-a_{KI\mu}\,a_{IJ\nu}\right).
\end{align*}
In a more explicit form, $\LC_{3,\text{KG}}$ is thus given by
\begin{align*}
\LC_{3,\text{KG}}&=\left(\pfrac{\phi_{I}^{*}}{x^{\alpha}}+
ig\,\phi_{J}^{*}\,a_{JI\alpha}\right)
\left(\pfrac{\phi_{I}}{x_{\alpha}}-ig\,a_{IJ}^{\alpha}\phi_{J}\right)-
m^{2}\,\phi_{I}^{*}\phi_{I}-\quarter p_{JK}^{\alpha\beta}\,p_{KJ\alpha\beta}
\end{align*}
The expressions in parentheses implement the ``minimum coupling rule,'' replacing ordinary derivatives by gauge-covariant derivatives.
Here they also give the field-theoretic canonical polymomenta.
They should not be identified with the mechanical momenta of a particle.
By inserting $\LC_{3,\text{KG}}$ into the Euler-Lagrange equations, and $\HC_{3,\text{KG}}$ into the canonical equations, we may convince ourselves that the emerging field equations for $\phi_{I}^{*}$, $\phi_{I}$, and $\ba_{JK}$ agree.
This means that $\HC_{3,\text{KG}}$ and $\LC_{3,\text{KG}}$ describe the {\em same physical system}.
\subsubsection{Dirac system Hamiltonian}
The generalized Dirac Hamiltonian~(\ref{hd-dirac}) describing $N$ spin-$\onehalf$ fields, each of them being associated with the same mass $m$,
\begin{equation*}
\HC_{\text{D}}=\left(\bar{\pi}_{I}^{\alpha}-\frac{i}{2}\bar{\psi}_{I}\gamma^{\alpha}
\right)\frac{3m^{*}\tau_{\alpha\beta}}{i}\left(\pi_{I}^{\beta}+\frac{i}{2}\gamma^{\beta}
\psi_{I}\right)+m\,\bar{\psi}_{I}\psi_{I},\quad\tau_{\mu\alpha}\sigma^{\alpha\nu}=
\delta_{\mu}^{\nu}\,\Eins
\end{equation*}
is form-invariant under global gauge transformations~(\ref{pointtra-rules}) since
\begin{align*}
\HC_{\text{D}}^{\prime}&=\left(\bar{\Pi}_{K}^{\alpha}-
\frac{i}{2}\bar{\Psi}_{K}\gamma^{\alpha}
\right)\frac{3m^{*}\tau_{\alpha\beta}}{i}\,\underbrace{u_{KI}u_{IJ}^{*}}_{=\delta_{KJ}}
\left(\Pi_{J}^{\beta}+\frac{i}{2}\gamma^{\beta}
\Psi_{J}\right)+m\,\bar{\Psi}_{K}\underbrace{u_{KI}u_{IJ}^{*}}_{=\delta_{KJ}}\Psi_{J}\\
&=\left(\bar{\Pi}_{K}^{\alpha}-\frac{i}{2}\bar{\Psi}_{K}\gamma^{\alpha}
\right)\frac{3m^{*}\tau_{\alpha\beta}}{i}\left(\Pi_{K}^{\beta}+\frac{i}{2}\gamma^{\beta}
\Psi_{K}\right)+m\,\bar{\Psi}_{K}\Psi_{K}.
\end{align*}
Again, the corresponding locally gauge-invariant Hamiltonian $\HC_{3,\text{D}}$ is found by adding the gauge Hamiltonian $\HC_{\text{g}}$ from Eq.~(\ref{H-g2})
\begin{align}
\HC_{3,\text{D}}&=\left(\bar{\pi}_{I}^{\alpha}-\frac{i}{2}\bar{\psi}_{I}\gamma^{\alpha}
\right)\frac{3m^{*}\tau_{\alpha\beta}}{i}\left(\pi_{I}^{\beta}+\frac{i}{2}\gamma^{\beta}
\psi_{I}\right)+m\,\bar{\psi}_{I}\psi_{I}\nonumber\\
&\quad\mbox{}-\quarter p_{JK}^{\alpha\beta}\,p_{KJ\alpha\beta}+
ig\left(\bar{\pi}_{K}^{\alpha}\,\psi_{J}-\bar{\psi}_{K}\,
\pi_{J}^{\alpha}+p_{JI}^{\alpha\beta}\,a_{IK\beta}\right)a_{KJ\alpha}.
\label{hd-final}
\end{align}
The correlation of the canonical momenta $\bar{\pi}_{I}^{\mu},\pi_{I}^{\mu}$ with the base fields $\bar{\psi}_{I},\psi_{I}$ and their derivatives follows again from first canonical equation for $\HC_{3,\text{D}}$
\begin{align}
\pfrac{\psi_{I}}{x^{\mu}}=\pfrac{\HC_{3,\text{D}}}{\bar{\pi}_{I}^{\mu}}&=
\frac{3m^{*}\tau_{\mu\beta}}{i}\left(\pi_{I}^{\beta}+\frac{i}{2}\gamma^{\beta}\psi_{I}\right)+
ig\,a_{IJ\mu}\psi_{J}\nonumber\\
\pfrac{\bar{\psi}_{I}}{x^{\mu}}=\pfrac{\HC_{3,\text{D}}}{\pi_{I}^{\mu}}&=
\left(\bar{\pi}_{I}^{\alpha}-\frac{i}{2}\bar{\psi}_{I}\gamma^{\alpha}
\right)\frac{3m^{*}\tau_{\alpha\mu}}{i}-ig\,\bar{\psi}_{J}\,a_{JI\mu}.
\label{pi-phip-dirac}
\end{align}
Inserting $\partial\psi_{I}/\partial x^{\mu}$ and $\partial\bar{\psi}_{I}/\partial x^{\mu}$ into Eq.~(\ref{general-invariant-lagrangian}), we encounter the related {\em locally\/} gauge-invariant Lagrangian $\LC_{3,\text{D}}$ in the intermediate form
\begin{equation}\label{ld-intermediate}
\LC_{3,\text{D}}=-\quarter p_{JK}^{\alpha\beta}\,p_{KJ\alpha\beta}+
\bar{\pi}_{I}^{\alpha}\frac{3m^{*}\tau_{\alpha\beta}}{i}\pi_{I}^{\beta}-
\left(m-m^{*}\right)\bar{\psi}_{I}\psi_{I},
\end{equation}
with the momenta $\bar{\pi}_{I}^{\alpha},\pi_{I}^{\beta}$ determined by Eqs.~(\ref{pi-phip-dirac}).
We can finally eliminate the momenta of the base fields in order to express $\LC_{3,\text{D}}$ completely in Lagrangian variables.
To this end, we solve Eqs.~(\ref{pi-phip-dirac}) for the momenta
\begin{align*}
\frac{3m^{*}\tau_{\alpha\beta}}{i}\pi_{I}^{\beta}&=\pfrac{\psi_{I}}{x^{\alpha}}-
ig\,a_{IK\alpha}\psi_{K}+\frac{im^{*}}{2}\gamma_{\alpha}\psi_{I}\\
\bar{\pi}_{I}^{\alpha}&=\left(\pfrac{\bar{\psi}_{I}}{x^{\beta}}+
ig\,\bar{\psi}_{J}a_{JI\beta}-\frac{im^{*}}{2}\bar{\psi}_{I}
\gamma_{\beta}\right)\frac{i\sigma^{\beta\alpha}}{3m^{*}}.
\end{align*}
Then
\begin{align*}
&\bar{\pi}_{I}^{\alpha}\frac{3m^{*}\tau_{\alpha\beta}}{i}\pi_{I}^{\beta}\\
&\,\,=\left(\pfrac{\bar{\psi}_{I}}{x^{\alpha}}+ig\,\bar{\psi}_{J}a_{JI\alpha}-
\frac{im^{*}}{2}\bar{\psi}_{I}\gamma_{\alpha}\right)\frac{i\sigma^{\alpha\beta}}{3m^{*}}\left(
\pfrac{\psi_{I}}{x^{\beta}}-ig\,a_{IK\beta}\psi_{K}+\frac{im^{*}}{2}\gamma_{\beta}\psi_{I}\right).
\end{align*}
Inserting this expression into~(\ref{ld-intermediate}) yields the final form of the locally gauge-invariant Dirac Lagrangian
\begin{align*}
\LC_{3,\text{D}}&=
\left(\pfrac{\bar{\psi}_{I}}{x^{\alpha}}+ig\,\bar{\psi}_{J}a_{JI\alpha}-
\frac{im^{*}}{2}\bar{\psi}_{I}\gamma_{\alpha}\right)\frac{i\sigma^{\alpha\beta}}{3m^{*}}\left(
\pfrac{\psi_{I}}{x^{\beta}}-ig\,a_{IK\beta}\psi_{K}+\frac{im^{*}}{2}\gamma_{\beta}\psi_{I}\right)\\
&\quad\mbox{}-\quarter p_{JK}^{\alpha\beta}p_{KJ\alpha\beta}-\left(m-m^{*}\right)\bar{\psi}_{I}\psi_{I}.
\end{align*}
After expanding, this Lagrangian writes equivalently
\begin{align}
\LC_{3,\text{D}}&=\frac{i}{2}\bar{\psi}_{I}\gamma^{\alpha}
\left(\pfrac{\psi_{I}}{x^{\alpha}}-ig\;a_{IK\alpha}\psi_{K}\right)-
\frac{i}{2}\left(\pfrac{\bar{\psi}_{I}}{x^{\alpha}}+
ig\,\bar{\psi}_{J}a_{JI\alpha}\right)\gamma^{\alpha}\psi_{I}-
m\,\bar{\psi}_{I}\psi_{I}\nonumber\\
&\quad\mbox{}+\left(\pfrac{\bar{\psi}_{I}}{x^{\alpha}}+
ig\,\bar{\psi}_{J}a_{JI\alpha}\right)\frac{i\sigma^{\alpha\beta}}{3m^{*}}
\left(\pfrac{\psi_{I}}{x^{\beta}}-ig\;a_{IK\beta}\psi_{K}\right)-
\quarter p_{JK}^{\alpha\beta}p_{KJ\alpha\beta}.\label{ld-final}
\end{align}
The sums in parentheses can be regarded as a generalized ``minimum coupling rule'' for the actual case of a Dirac Lagrangian describing an $N$-tuple of spinors $\psi_{I}$.
This also applies for the term involving $\sigma^{\alpha\beta}$ in Eq.~(\ref{ld-final}) that emerges {\em in addition\/} to the conventional gauge-invariant Lagrangian if we start from the ``regularized'' Lagrangian from Eq.~(\ref{ld-dirac-regular}).
This term is easily shown to be separately form-invariant under the combined local gauge transformation that is defined by Eqs.~(\ref{general-pointtra}) and (\ref{gauge-tra1}).
The bilinear $\bar{\psi}_{J}\sigma^{\alpha\beta}\psi_{I}$ transforms as a Lorentz $(2,0)$-tensor.
It is not itself a Lorentz scalar.
The fully contracted term in the Lagrangian is Lorentz-invariant.
Physically, the term describes Pauli-coupling of the $N$-tuple of fermions $\psi_{I}$ with the matrix of bosonic $4$-vector gauge fields $a_{IK\mu}$.

The $\bp_{KJ}$ stand for the combinations of the Lagrangian dynamical variables of the gauge fields from Eq.~(\ref{can-momentum-gf}) for the chosen gauge kinetic term
\begin{equation*}
p_{KJ\alpha\beta}=\pfrac{a_{KJ\beta}}{x^{\alpha}}-\pfrac{a_{KJ\alpha}}{x^{\beta}}+
ig\left(a_{KI\beta}\,a_{IJ\alpha}-a_{KI\alpha}\,a_{IJ\beta}\right).
\end{equation*}
In order to set up the Euler-Lagrange equations for the locally gauge-invariant Lagrangian $\LC_{3,\text{D}}$ from Eq.~(\ref{ld-final}), we first calculate the derivatives
\begin{align*}
\pfrac{}{x^{\alpha}}\pfrac{\LC_{3,\text{D}}}{\left(\partial_{\alpha}\bar{\psi}_{I}\right)}&=
-\frac{i}{2}\gamma^{\alpha}\pfrac{\psi_{I}}{x^{\alpha}}+\frac{i\sigma^{\alpha\beta}}{3m^{*}}\left(
\cancel{\pfrac{^2\psi_{I}}{x^{\alpha}\partial x^{\beta}}}-ig\pfrac{a_{IK\beta}}{x^{\alpha}}
\psi_{K}-ig\,a_{IK\beta}\pfrac{\psi_{K}}{x^{\alpha}}\right)\\
\pfrac{\LC_{3,\text{D}}}{\bar{\psi}_{I}}&=
\frac{i}{2}\gamma^{\alpha}\pfrac{\psi_{I}}{x^{\alpha}}-
m\psi_{I}+g\,a_{IK\alpha}\gamma^{\alpha}\psi_{K}\\
&\quad\mbox{}+\frac{i\sigma^{\alpha\beta}}{3m^{*}}\,ig\left(a_{IK\alpha}
\pfrac{\psi_{K}}{x^{\beta}}-ig\,a_{IJ\alpha}a_{JK\beta}\psi_{K}\right)
\end{align*}
and
\begin{align*}
\pfrac{}{x^{\beta}}\pfrac{\LC_{3,\text{D}}}{\left(\partial_{\beta}\psi_{I}\right)}&=
\frac{i}{2}\pfrac{\bar{\psi}_{I}}{x^{\beta}}\gamma^{\beta}+\left(
\cancel{\pfrac{^2\bar{\psi}_{I}}{x^{\alpha}\partial x^{\beta}}}+ig\,
\pfrac{\bar{\psi}_{K}}{x^{\beta}}a_{KI\alpha}+
ig\,\bar{\psi}_{K}\pfrac{a_{KI\alpha}}{x^{\beta}}\right)\frac{i\sigma^{\alpha\beta}}{3m^{*}}\\
\pfrac{\LC_{3,\text{D}}}{\psi_{I}}&=-\frac{i}{2}\pfrac{\bar{\psi}_{I}}{x^{\alpha}}\gamma^{\alpha}-
m\bar{\psi}_{I}+g\,\bar{\psi}_{K}\gamma^{\alpha}a_{KI\alpha}\\
&\quad\mbox{}+ig\left(\pfrac{\bar{\psi}_{K}}{x^{\beta}}a_{KI\alpha}-
ig\,\bar{\psi}_{K}a_{KJ\alpha}a_{JI\beta}\right)\frac{i\sigma^{\alpha\beta}}{3m^{*}}.
\end{align*}
The second derivative terms drop out due to the skew-symmetry of $\sigma^{\alpha\beta}$.
The Euler-Lagrange equations thus finally emerge as
\begin{align}
i\gamma^{\alpha}\pfrac{\psi_{I}}{x^{\alpha}}+g
\gamma^{\alpha}a_{IK\alpha}\psi_{K}-m\,\psi_{I}-
\frac{g}{6m^{*}}p_{IK\alpha\beta}\sigma^{\alpha\beta}\psi_{K}&=0\nonumber\\
i\pfrac{\bar{\psi}_{I}}{x^{\alpha}}\gamma^{\alpha}-
g\,\bar{\psi}_{K}a_{KI\alpha}\gamma^{\alpha}+m\,\bar{\psi}_{I}+
\frac{g}{6m^{*}}\bar{\psi}_{K}\sigma^{\alpha\beta}p_{KI\alpha\beta}&=0.
\label{el-dirac-gi}
\end{align}
We observe that our gauge-invariant Dirac equation contains an additional term that is proportional to $p_{IK\alpha\beta}$, hence to the canonical momenta of the gauge fields $a_{IK\alpha}$.
This term is gauge covariant, transforming in the same way as the other terms of the spinor equation.
We thus encounter the description of the coupling of the anomalous magnetic moments of the fermions to the gauge bosons, i.e., a spin-gauge field coupling.

For a single spinor $\psi$ of mass $m\ne0$, hence for the U$(1)$ gauge group, one may choose the particular model $m^{*}=m$.
This identification is not required by gauge symmetry.
For this choice, the locally gauge-covariant Dirac equation reduces to
\begin{equation*}
i\gamma^{\alpha}\pfrac{\psi}{x^{\alpha}}+g\,\gamma^{\alpha}a_{\alpha}\psi-
m\,\psi-\frac{\mu}{3}\left(\pfrac{a_{\beta}}{x^{\alpha}}-
\pfrac{a_{\alpha}}{x^{\beta}}\right)\sigma^{\alpha\beta}\psi=0,
\end{equation*}
with $\mu=g/2m$ the particle's magneton.
The equation is obviously invariant under the combined gauge transformation of base and gauge fields
\begin{equation*}
a_{\mu}(\bx)\mapsto A_{\mu}(\bx)=a_{\mu}(\bx)+
\frac{1}{g}\pfrac{\Lambda(\bx)}{x^{\mu}},\qquad
\psi(\bx)\mapsto\Psi(\bx)=\psi(\bx)\,e^{i\Lambda(\bx)},
\end{equation*}
with the spin-gauge field coupling term transforming covariantly with the same phase as the other spinor terms.
Here, the additional term corresponds to a coupling of the electromagnetic field with the spin-induced magnetic moment of the fermion represented by $\psi$, commonly referred to as ``Pauli-coupling'' term.
Pauli interaction thus emerges when the present construction is applied to the particular regular Dirac representative in Eq.~(\ref{ld-dirac-regular}).
Equivalently, it is obtained by applying the minimum coupling rule to that Lagrangian.
It is not required by local gauge symmetry alone.
A separately gauge-invariant Pauli term with an independent coefficient may also be added, including one that cancels this contribution.
\subsubsection{\label{sec:pauli-comparison}Comparison with Pauli's amended Lagrangian}
In this context, we remark that the Pauli-coupling term in the field equations~(\ref{el-dirac-gi}) equally follows from the amended Dirac Lagrangian
\begin{align}
\LC_{3,\text{Pauli}}&=\frac{i}{2}\bar{\psi}_{I}\gamma^{\alpha}
\left(\pfrac{\psi_{I}}{x^{\alpha}}-ig\;a_{IK\alpha}\psi_{K}\right)-
\frac{i}{2}\left(\pfrac{\bar{\psi}_{I}}{x^{\alpha}}+
ig\,\bar{\psi}_{J}a_{JI\alpha}\right)\gamma^{\alpha}\psi_{I}-
m\,\bar{\psi}_{I}\psi_{I}\nonumber\\
&\quad\mbox{}-\onehalf\ell\,\bar{\psi}_{J}p_{JK\alpha\beta}
\sigma^{\alpha\beta}\psi_{K}-\quarter p_{JK}^{\alpha\beta}p_{KJ\alpha\beta}
\label{ld-pauli}
\end{align}
if we identify the coupling constant as
\begin{equation*}
\ell=\frac{g}{3m^{*}},\qquad [\ell]=[L].
\end{equation*}
For the particular choice $m^{*}=m$, this gives $\ell=g/(3m)$.
The addition of such a term was proposed by Pauli\cite{pauli}.
One fixed sign in the Lagrangian gives both the equation for $\psi_{I}$ and the equation for its Dirac adjoint $\bar\psi_{I}$.
The latter is not the charge-conjugate spinor.
No change in the sign of $\ell$ is made when varying the adjoint field.

The relation between Eqs.~(\ref{ld-final}) and~(\ref{ld-pauli}) can be established directly by the product rule.
Write
\begin{align*}
(D_{\alpha}\psi)_{I}&=\pfrac{\psi_{I}}{x^{\alpha}}
-ig\,a_{IK\alpha}\psi_{K},\\
(D_{\alpha}\bar\psi)_{I}&=\pfrac{\bar\psi_{I}}{x^{\alpha}}
+ig\,\bar\psi_{K}a_{KI\alpha}.
\end{align*}
For the field strength $p_{IK\alpha\beta}$ defined by Eq.~(\ref{can-momentum-gf}),
\begin{equation*}
[D_{\alpha},D_{\beta}]\psi_{I}
=-ig\,p_{IK\alpha\beta}\psi_{K},\qquad
\sigma^{\alpha\beta}(D_{\alpha}D_{\beta}\psi)_{I}
=-\frac{ig}{2}\sigma^{\alpha\beta}p_{IK\alpha\beta}\psi_{K}.
\end{equation*}
Since the contracted internal indices form a gauge singlet,
\begin{equation*}
\pfrac{}{x^{\alpha}}\left[
\bar\psi_{I}\sigma^{\alpha\beta}(D_{\beta}\psi)_{I}\right]
=(D_{\alpha}\bar\psi)_{I}\sigma^{\alpha\beta}(D_{\beta}\psi)_{I}
+\bar\psi_{I}\sigma^{\alpha\beta}(D_{\alpha}D_{\beta}\psi)_{I}.
\end{equation*}
Consequently, for constant $m^{*}$,
\begin{align*}
\frac{i}{3m^{*}}(D_{\alpha}\bar\psi)_{I}
\sigma^{\alpha\beta}(D_{\beta}\psi)_{I}
&=\pfrac{}{x^{\alpha}}\left[
\frac{i}{3m^{*}}\bar\psi_{I}\sigma^{\alpha\beta}
(D_{\beta}\psi)_{I}\right]\\
&\quad-\frac{g}{6m^{*}}\bar\psi_{I}\sigma^{\alpha\beta}
p_{IK\alpha\beta}\psi_{K}.
\end{align*}
This off-shell identity also fixes the Pauli signs in Eq.~(\ref{el-dirac-gi}).
It proves the non-Abelian relation
\begin{equation*}
\LC_{3,\mathrm{D}}=\LC_{3,\mathrm{Pauli}}
+\pfrac{B^{\alpha}}{x^{\alpha}},\qquad
B^{\alpha}=\frac{i}{3m^{*}}\bar\psi_{I}\sigma^{\alpha\beta}
(D_{\beta}\psi)_{I}.
\end{equation*}
The two forms therefore give identical bulk Euler-Lagrange equations for the spinors and the gauge fields, for compactly supported variations or boundary conditions that remove the variation of the boundary integral.

With the corrected sign and normalization, the expressions used for comparison may be retained as
\begin{align*}
\LC_{\text{int,Pauli}}&=-\ell\,\bar{\psi}_{I}\left(
\pfrac{a_{IJ\beta}}{x^{\alpha}}+ig\,a_{IK\beta}\,a_{KJ\alpha}\right)
\sigma^{\alpha\beta}\psi_{J},\\
\LC_{\text{int}}&=-\frac{\ell}{ig}\left(\pfrac{\bar{\psi}_{I}}{x^{\alpha}}+
ig\,\bar{\psi}_{J}a_{JI\alpha}\right)\sigma^{\alpha\beta}
\left(\pfrac{\psi_{I}}{x^{\beta}}-ig\,a_{IK\beta}\psi_{K}\right).
\end{align*}
The second expression is the covariantized regularizing term, including its free-theory boundary contribution, rather than a pure interaction term.
It is produced by minimal coupling of the chosen regular Dirac representative.
The first expression is the usual non-minimal Pauli operator.
Their difference is precisely the divergence displayed above.

Although their bulk equations agree, the two Lagrangian representatives need not have the same canonical momenta.
For example, in the Abelian case Eq.~(\ref{ld-pauli}) gives
\begin{equation*}
\pfrac{\LC_{3,\mathrm{Pauli}}}{(\partial_{\nu}a_{\mu})}
=p^{\mu\nu}+\ell\bar\psi\sigma^{\mu\nu}\psi
=p^{\mu\nu}+\frac{g}{3m^{*}}\bar\psi\sigma^{\mu\nu}\psi.
\end{equation*}
Here $p^{\mu\nu}=\partial^{\mu}a^{\nu}-\partial^{\nu}a^{\mu}$ is the field-strength abbreviation, not the gauge momentum of the Pauli representative.
The distinction concerns the canonical representation and possible boundary contributions.
Different-looking interaction expressions do not establish inequivalent bulk physics or select one representative as physically preferable.

Finally, the parameter $m^{*}$ has a different role in the free and interacting theories.
In the free theory it multiplies a null-Lagrangian contribution.
After covariantization, its value determines the Pauli coupling $\ell=g/(3m^{*})$ and is therefore a physical model parameter, not a gauge choice.
\section{Conclusions}
With the present paper, we have worked out a consistent local inertial frame description of the canonical formalism in the realm of covariant Hamiltonian field theory.
On that basis, Noether's theorem has been formulated within the admissible class of canonical transformations.
The corresponding local U$(N)$ construction determines the minimal matter coupling, while the quadratic gauge kinetic term is an additional invariant dynamical choice.
The gauge-field momentum sector must be treated with the first-order action and its antisymmetry condition.
Applied to the selected regular Dirac representative, the construction produces a Dirac-Pauli theory with coupling controlled by the independent constant $m^{*}$.
The covariantized regular Lagrangian and the matched Dirac-Pauli Lagrangian differ by a total divergence.
No inequivalence of their bulk field equations or physical predictions follows from their different canonical representations.
\begin{acknowledgement}
To the memory of my (J.S.) colleague and friend Dr.~Claus~Riedel (GSI), who contributed vitally to this work.
Furthermore, the authors are indebted to Prof.~Dr.~Dr.~hc.~mult.\ Walter Greiner from the {\em Frankfurt Institute of Advanced Studies\/} (FIAS) for his long-standing hospitality, his critical comments and encouragement.
\end{acknowledgement}


\begin{thebibliography}{99}
\bibitem{dedonder} Th.~De~Donder, {\em Th\'eorie Invariantive Du Calcul des Variations}, (Gaulthier-Villars \& Cie., Paris, 1930).
\bibitem{weyl} H.~Weyl, Geodesic Fields in the Calculus of Variation for Multiple Integrals, in {\em Annals of Mathematics} {\bf 36} 607 (1935).
\bibitem{saletan} See, for instance, J.~V.~Jos\'e and E.~J.~Saletan, {\em Classical Dynamics}, Cambridge University Press, Cambridge, 1998.
\bibitem{greiner} W.~Greiner, B.~M\"uller, and J.~Rafelski, {\em Quantum Electrodynamics of Strong Fields}, (Springer-Verlag, Berlin, 1985).
\bibitem{gasi} S.~Gasiorowicz, {\em Elementary particle physics}, (Wiley, New York, 1966).
\bibitem{vonRieth} J.~von Rieth, The Hamilton-Jacobi theory of De~Donder and Weyl applied to some relativistic field theories, in {\em J.~Math.~Phys.} {\bf 25}, 1102 (1984).
\bibitem{noether} E.~Noether, Invariante Variationsprobleme, Nachr.~Ges.~Wiss.~G\"ottingen, Math.-Phys.\ Kl., 235--257 (1918).
\bibitem{pauli} W.~Pauli, {\em Rev.~Mod.~Phys.\/} {\bf 13}, 203--232 (1941).
\end{thebibliography}
\end{document}